\documentclass[twocolumn, trackchanges]{aastex62}
\received{2019 May 30}
\revised{2019 September 15}
\accepted{2019 September 25}

\shorttitle{A catalog of galaxies in the direction of the Perseus cluster}
\shortauthors{Wittmann et al.}

\begin{document}

\title{A CATALOG OF GALAXIES IN THE DIRECTION OF THE PERSEUS CLUSTER}

\correspondingauthor{Carolin Wittmann}
\email{carolin@dwarfgalaxies.net}

\author[0000-0003-2588-8862]{Carolin Wittmann}
\affiliation{Astronomisches Rechen-Institut, Zentrum f\"ur Astronomie der Universit\"at Heidelberg, M\"onchhofstra{\ss}e 12-14, 69120 Heidelberg, Germany}

\author[0000-0002-4460-9892]{Ralf Kotulla}
\affiliation{Department of Astronomy, University of Wisconsin at Madison, 475 North Charter Street, Madison, WI 53706-1582, USA}

\author[0000-0002-6807-5856]{Thorsten Lisker}
\affiliation{Astronomisches Rechen-Institut, Zentrum f\"ur Astronomie der Universit\"at Heidelberg, M\"onchhofstra{\ss}e 12-14, 69120 Heidelberg, Germany}

\author[0000-0002-1891-3794]{Eva K. Grebel}
\affiliation{Astronomisches Rechen-Institut, Zentrum f\"ur Astronomie der Universit\"at Heidelberg, M\"onchhofstra{\ss}e 12-14, 69120 Heidelberg, Germany}

\author[0000-0003-1949-7638]{Christopher J. Conselice}
\affiliation{School of Physics and Astronomy, University of Nottingham, Nottingham NG7 2RD, UK}

\author{Joachim Janz}
\affiliation{Astronomy Research Unit, University of Oulu, Pentti Kaiteran katu 1, 90014 Oulu, Finland}
\affiliation{Finnish Centre of Astronomy with ESO (FINCA), University of Turku, Väisäläntie 20, 21500 Piikkiö, Finland}

\author[0000-0001-5703-7531]{Samantha J. Penny}
\affiliation{Institute of Cosmology and Gravitation, University of Portsmouth, Burnaby Road, Portsmouth PO1 3FX, UK}




\begin{abstract}
We present a catalog of $5437$ morphologically classified sources in the direction of the Perseus galaxy cluster core, among  them $496$ early-type low-mass galaxy candidates. The catalog is primarily based on $V$-band imaging data acquired with the {\it William Herschel Telescope}, which we used to conduct automated source detection and to derive photometry. We additionally reduced archival {\it Subaru} multiband imaging data in order to measure aperture colors and to perform a morphological classification, benefiting from 0.5 arcsec seeing conditions in the $r$-band data. Based on morphological and color properties, we extracted a sample of early-type low-mass galaxy candidates with absolute $V$-band magnitudes in the range of $-10$ to $-20$\,mag. In the color-magnitude diagram the galaxies are located where the red sequence for early-type cluster galaxies is expected, and they lie on the literature relation between absolute magnitude and S\'{e}rsic index. We classified the early-type dwarf candidates into nucleated and nonnucleated galaxies. For the faint candidates, we found a trend of increasing nucleation fraction toward brighter luminosity or higher surface brightness, similar to what is observed in other nearby galaxy clusters. We morphologically classified the remaining sources as likely background elliptical galaxies, late-type galaxies, edge-on disk galaxies, and likely merging systems and discussed the expected contamination fraction through non-early-type cluster galaxies in the magnitude-size surface brightness parameter space. Our catalog reaches its $50$\,per\,cent completeness limit at an absolute $V$-band luminosity of $-12$\,mag and a $V$-band surface brightness of $26$\,mag\,arcsec$^{-2}$. This makes it to the largest and deepest catalog with coherent coverage compared to previous imaging studies of the Perseus cluster.
\end{abstract}



\section{Introduction} \label{sec:sec1}

Galaxy clusters are the densest regions with the highest galaxy number densities in the universe. They contain a wide variety of stellar systems, ranging from the cluster central galaxy, which is typically the most luminous cluster member; over all kinds of early- and late-type galaxies from the giant to the dwarf scale; to very small stellar systems like compact elliptical galaxies, ultracompact dwarf galaxies, and globular clusters. 

Galaxy clusters continuously grow by accreting individual galaxies, as well as entire galaxy groups. At the same time, they transform or even destroy the newly arrived members. Ram pressure, exerted by the intracluster medium, leads to stripping and depletion of an infalling galaxy's gas reservoir \citep{Gunn1972}, preventing the fueling of further star formation and turning the galaxy red in color due to its aging stellar population. Tidal stripping or heating induced by the overall cluster potential or close passages of massive cluster galaxies may result in a perturbation or even the destruction of the stellar galaxy component \citep[e.g.,][]{Moore1998, Gnedin2003, Bialas2015, Smith2015}. These cluster environmental processes result in different properties for cluster and field galaxies and are thought to give rise to the observed relations between morphology and color with environmental density \citep{Dressler1980, Binggeli1987, Bamford2009}, where red non-star-forming early-type galaxies (ETGs) dominate in high-density regions, and blue star-forming late-type galaxies (LTGs) are preferentially located in lower-density environments.

Although dwarf galaxies only make up a minor fraction of the total luminosity and mass content of all cluster galaxies, they are the dominant cluster population by numbers. This makes them to ideal tracers of a cluster's assembly history. Signatures of their infall and accretion history will still be imprinted in their spatial and velocity distribution after several gigayears, since dwarf galaxies are insensitive to dynamical friction \citep{Vijayaraghavan2015}. Dwarf galaxies are also valuable indicators to study cluster environmental processes. Due to their low masses and shallow gravitational potential wells, dwarf galaxies are thought to be particularly sensitive to external influences \citep[e.g.,][]{Penny2014b}.

Pioneering work in cataloging and studying dwarf populations in galaxy clusters was carried out by \citet{Binggeli1985} in the Virgo cluster and \citet{Ferguson1989} in the Fornax cluster. Until today, these two environments constituted the best-studied nearby clusters in terms of their galaxy populations, from which a lot of insight was gained with regard to understanding galaxy evolution. Also, the more distant Coma galaxy cluster generated a lot of interest, leading to a well-studied galaxy population in an even more massive cluster environment compared to Virgo and Fornax \citep[e.g.,][]{Godwin1983, Carter2008, Michard2008}. Recent deep wide-field imaging surveys of Virgo (Next Generation Virgo Cluster Survey (NGVS); \citealt{Ferrarese2012}), Fornax (Fornax Deep Survey (FDS); \citealt{Venhola2018}; Next Generation Fornax Cluster Survey (NGFS); \citealt{Munoz2015}), and Coma \citep{Koda2015} pushed the detection limits to very faint luminosities and surface brightnesses, allowing one to detect dwarf galaxy members even in the regime of Local Group dwarf spheroidals.

The Perseus galaxy cluster is another nearby rich galaxy cluster at a distance of $\sim 70$\,Mpc. With a virial mass on the order of $8.5 \times 10^{14}\,M_\odot$ \citep{Mathews2006}; it is clearly more massive than Fornax and Virgo and by a factor of $\sim 1.5$ less massive than the more distant Coma cluster. There are signs that Perseus is a dynamically young environment, like Virgo, with indications of ongoing assembly \citep{Andreon1994, Brunzendorf1999}. The cluster has a variety of interesting properties, for example, the peculiar central galaxy NGC\,1275, very bright X-ray emission, and a high cluster velocity dispersion \citep[see][]{Conselice2002}. This makes Perseus stand out as another unique rich cluster environment in the nearby Universe with interesting prospects to study environmental influences on its galaxy population. Nevertheless, Perseus has not yet been studied in a similar detail as Virgo, Fornax, and Coma.

The Perseus cluster is partly covered by the Sloan Digital Sky Survey (SDSS; \citealt{Ahn2012}), but the data do not probe far into the dwarf galaxy luminosity and size regime at the distance of Perseus, due to the rather shallow depth and comparably low spatial resolution of the data. \citet{Brunzendorf1999} presented a catalog of 660 brighter galaxies, with $B_{25} < 19.5$\,mag, which they detected on a 10\,deg$^2$ area of the Perseus cluster based on digitized Schmidt plates. Deeper and higher-resolution imaging studies of a $0.3 \times 0.3$\,Mpc$^2$ region of the cluster core  were conducted by \citet{Conselice2002, Conselice2003}, based on multiband-imaging data acquired with the {\it WIYN} $3.5$\,m telescope, revealing a sample of $53$ dwarf galaxy candidates. About $30$ possible and confirmed dwarf cluster members, partly overlapping with the above sample, were studied in detail based on observations with the {\it Hubble Space Telescope (HST)} by \citet{deRijcke2009} and \citet{Penny2009}. A deep and wide-field survey conducted with the {\it William Herschel Telescope (WHT)} was presented in \citet{Wittmann2017}. Their observations cover $0.7 \times 0.7$\,Mpc$^2$ of the central cluster region and reach a $5\sigma$ $V$-band depth of $\sim 27$\,mag\,arcsec$^{-2}$, which allowed them to identify a population of $89$ faint low surface brightness galaxy candidates with mean effective $V$-band surface brightness fainter than $25$\,mag\,arcsec$^{-2}$.

In this study, we present a new catalog of sources in the direction of the Perseus cluster core based on the {\it WHT} $V$-band data described in \citet{Wittmann2017}. It is a collection of both likely cluster members and sources in the background, with our primary focus being extracting a sample of early-type low-mass galaxy candidates. We used the {\it WHT} $V$-band mosaic to perform an automated source detection with \textsc{SExtractor} \citep{Bertin1996}, which we tuned to be most efficient at detecting faint extended sources. The most reliable approach in discriminating cluster from background galaxies is based on radial velocity measurements. However, in the faint luminosity regime of dwarf galaxies, so far only a handful of sources have been spectroscopically confirmed in Perseus \citep{Penny2008}. We therefore performed a morphological analysis of all sources detected in our {\it WHT} $V$-band mosaic. Selecting early-type dwarf galaxies based on their morphological appearance has been demonstrated to be quite successful, e.g., by \citet{Binggeli1985} and \citet{Conselice2002}. However, the success of a correct morphological classification greatly depends on the spatial resolution of the imaging data in order to reveal the substructure of possible background galaxies. Therefore, we examined archival {\it Subaru} HyperSuprimeCam (HSC) data with very good seeing conditions that overlap with our {\it WHT} $V$-band footprint, allowing us to perform a very high quality morphological classification even for small and faint sources. 

This paper is structured as follows: In Section~\ref{sec:sec2} we characterize the data sets used for this study. In Section~\ref{sec:sec3} we outline the source detection and give a quantitative completeness estimate for the catalog. The photometric measurements are described in Section~\ref{sec:sec4}. In Section~\ref{sec:sec5} we explain the morphological classification. Our final catalog is presented in Section~\ref{sec:sec6}. We provide a discussion on the parameter distributions of the morphologically classified sources in Section~\ref{sec:sec7}, and perform a literature comparison in Section~\ref{sec:sec8}. We summarize the results of this study in Section~\ref{sec:sec9}.

Throughout the paper, we adopt a distance of $72.3$\,Mpc to the Perseus cluster and a scale of $20.32$\,kpc arcmin$^{-1}$ \citep[][using the `cosmology-corrected' quantities from the NASA/IPAC Extragalactic Database (NED), with $H_0 =  73.0$\,km\,s$^{-1}$\,Mpc$^{-1}$, $\Omega_{\mathrm{matter}} = 0.27$, $\Omega_{\mathrm{vacuum}} = 0.73$]{Struble1999}.

\section{Data} \label{sec:sec2}

\begin{figure*}
\centering
	\includegraphics[width=0.9\textwidth]{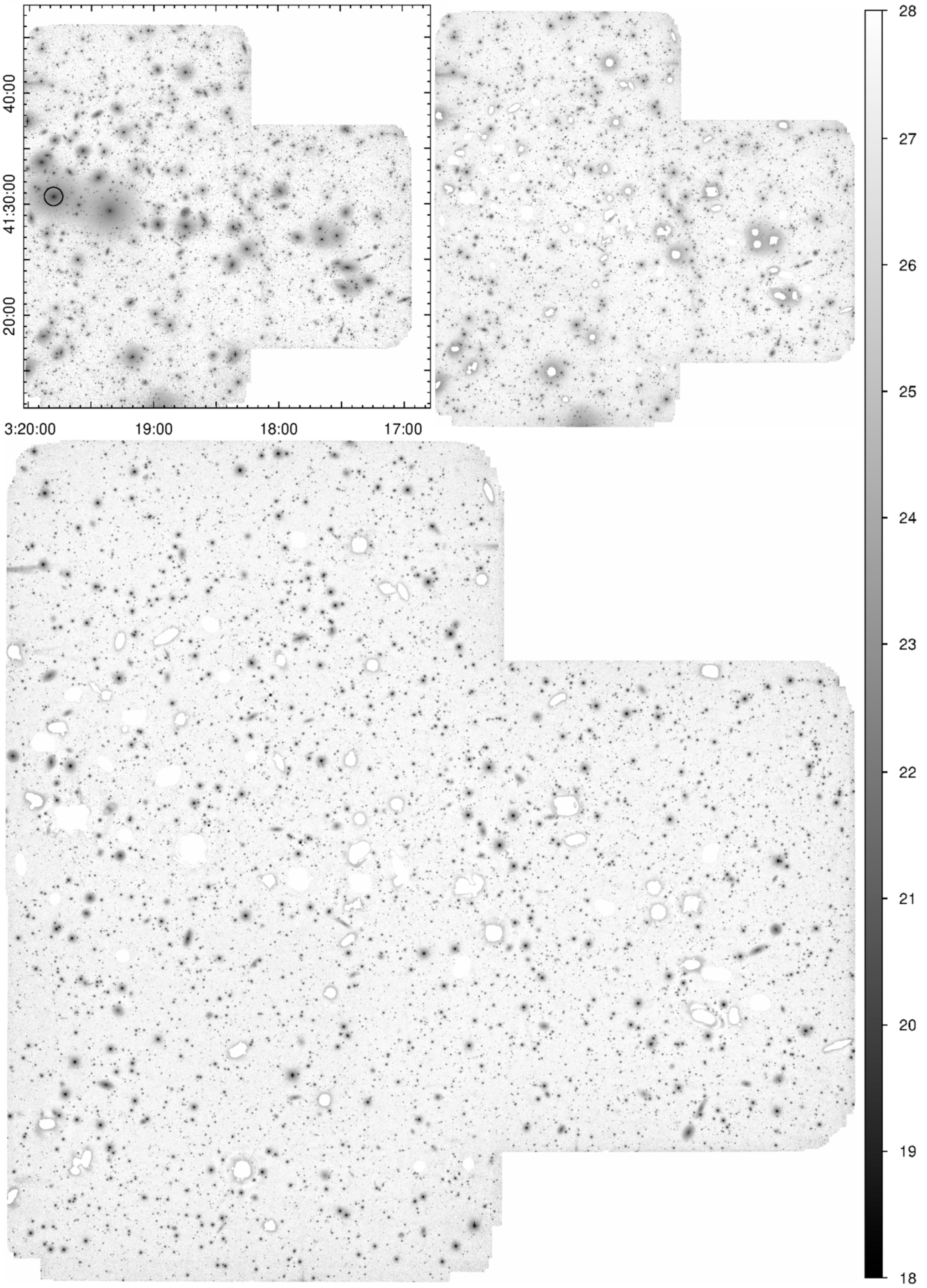}
    \caption{Top left: reduced {\it WHT} $V$-band mosaic of the Perseus cluster core. The $x$- and $y$-axes correspond to R.A. and decl., respectively. The black circle marks NGC\,1275. Top right: partly object-subtracted mosaic with overlaid mask indicating the excluded inner regions of extended high surface brightness sources. Bottom: mosaic with subtracted \textsc{SExtractor} background map (\texttt{BACK\_SIZE} $= 64$\,pixels) and the high surface brightness object mask used for source detection. The gray-scale bar indicates the surface brightness in mag\,arcsec$^{-2}$. North is up, and east is to the left.}
    \label{fig:fig1}
\end{figure*}

We performed the source detection and photometry in our $V$-band mosaic of the Perseus cluster core, which we presented and described in \citet[][see their Figure 1, and Figure~\ref{fig:fig1} in this paper]{Wittmann2017}. The mosaic is based on imaging data acquired with the Prime Focus Imaging Platform at the {\it WHT} (program 2012B/045; PI: T. Lisker). It covers a region of $0.27$\,deg$^2$, which corresponds to $0.41$\,Mpc$^2$ at the distance of Perseus and extends to a clustercentric distance of $0.57$\,\degr ($\widehat{=}\,0.70$\,Mpc). The mosaic is characterized by an image depth of $27$\,mag\,arcsec$^{-2}$ in the $V$-band at a signal-to-noise ratio ($S/N$) of $1$ per pixel, with a pixel scale of $0.237$\,arcsec\,pixel$^{-1}$. The average seeing full width at half-maximum (FWHM) corresponds to $0.9$\,arcsec. 

The {\it WHT} data constitute our primary data set that we used to establish the catalog. When starting to work on the catalog, we had the fully reduced {\it WHT} $V$-band mosaic already in hand, and we were convinced that the data would be well suited for automated source detection and photometry based on our experience in \citet{Wittmann2017}, where we used the data for visual detection and photometry of faint low surface brightness galaxy candidates. We additionally reduced archival {\it Subaru} HSC data in the $g$-, $r$-, and $z$-bands, which served as an auxiliary data set to improve the quality of the catalog. This allowed us to benefit from the 0.5\,arcsec seeing conditions in the {\it Subaru} $r$-band data for the morphological analysis, as well as to obtain color information by deriving $g-r$ and $r-z$ aperture colors.

\begin{deluxetable}{lcccc}
\tablecaption{Summary of Image Statistics for the Subaru HSC Data\label{tab:hsc}}
\tablecolumns{5}
\tablewidth{\textwidth}
\tablehead{
\colhead{Filter} & \colhead{Exposure} & \colhead{FWHM} & \colhead{Limiting} & \colhead{Surface}\\
 & \colhead{Time (s)} & \colhead{(arcsec)} & \colhead{Magnitude\tablenotemark{a}} & \colhead{Brightness\tablenotemark{b}}
}
\startdata
         HSC-$g$ & 23,760 & 0.64 & 27.6 & 28.8 \\
         HSC-$r$ & 4320 & 0.46 & 26.5 & 27.6 \\
         HSC-$z$ & 2880 & 0.60 & 24.9 & 25.9 \\
\enddata
    \tablenotetext{a}{Magnitude for point sources detected at $\geq 10\sigma$ (i.e. with photometric uncertainties $<0.1$ mag).}
    \tablenotetext{b}{Limiting surface brightness in mag\,arcsec$^{-2}$ per pixel (pixel size 0.237\,arcsec).}
\end{deluxetable}

A full description of the reduction details for the {\it Subaru} data set is presented in a forthcoming paper by Kotulla et al., so we only present a short overview to provide critical information. The HSC dataset was compiled from a number of mostly overlapping observations taken under two different proposals in 2014 September and November, obtained from the Subaru-Mitaka-Okayama-Kiso Archive  \citep[SMOKA,][]{smoka2002}. Data reduction, including bias, dark, and flat-field correction, followed by astrometric and photometric calibration, was performed using the HSC pipeline (Version 5.4). All detrended and calibrated images were then aligned and coadded into a deep stack. To measure aperture colors, we coaligned all bandpasses to the same pixel grid of the {\it WHT} data, allowing us to reuse masks and photometry apertures consistently across all bands and thus limiting systematic uncertainties and biases. This was done using \textsc{SWarp} \citep{Bertin2002} to match pixel scale, pointing, and field of view. Some statistics describing these observations are summarized in Table \ref{tab:hsc}. For the morphological analysis, we used the $r$-band data with the native HSC pixel scale of $0.168$\,arcsec\,pixel$^{-1}$. 

We publicly release the mosaics used in this study through the German Astrophysical Virtual Observatory (GAVO); see Section~\ref{sec:sec6}.

\section{The source catalog} \label{sec:sec3}

\subsection{Source Detection} \label{sec:sec3.1}

We established a catalog of sources in the direction of Perseus based on detection with \textsc{SExtractor}\footnote{\textsc{SExtractor} version $2.8.6$.} on our {\it WHT} $V$-band mosaic. Prior to detection, we fitted and subtracted the light profiles of most of the bright galaxies and stellar halos using \textsc{iraf}\footnote{\textsc{iraf} is distributed by the National Optical Astronomy Observatory, which is operated by the Association of Universities for Research in Astronomy (AURA) under a cooperative agreement with the National Science Foundation.} \emph{ellipse}. We additionally excluded the inner regions of all subtracted sources, since they often show pronounced residuals, as well as the bright centers of remaining sources that were not subtracted. We therefore masked all regions that were detected with \textsc{SExtractor} in the original Perseus mosaic (where no sources were subtracted) when requiring a detection threshold above $5\,\sigma$ and more than $10,000$ connected pixels. We note that this results in a bright $V$-band magnitude limit of $-20$\,mag for our catalog and a reduced detection efficiency for the larger sources with absolute $V$-band magnitudes brighter than $-19$\,mag (see Figure~\ref{fig:fig4} and Section~\ref{sec:sec3.3}). The masked regions in the partly object-subtracted {\it WHT} mosaic are indicated in Figure~\ref{fig:fig1} (top right and bottom panels).

We used a set of model galaxies inserted into our data to tailor the \textsc{SExtractor} parameter configuration especially to the detection of faint galaxies. We generated a set of $69$ model galaxies spanning the parameter range of faint low-mass galaxies from compact elliptical to faint low surface brightness galaxies with absolute $V$-band magnitudes $M_{V,0} = -10$ to $-19$\,mag ($m_{V,0} = 24.3 - 15.3$\,mag), half-light radii $r_{50} = 0.2 - 7.8$\,kpc ($r_{50} = 0.6 - 23.0$\,arcsec), and effective surface brightnesses $\langle\mu_{V,0}\rangle_{50} = 16 - 27$\,mag\,arcsec$^{-2}$ at the distance of Perseus. In the following, we denote extinction-corrected magnitudes by adding the subscript "$0$". For the model galaxies, we assumed an extinction of $A_V = 0.5$\,mag, which is the average value for the region covered by the {\it WHT} mosaic footprint. For all nonartificial sources, we obtained the extinction values at the position of the respective source (see Section~\ref{sec:sec4.3}, and \ref{sec:sec4.4}).

We realized all model galaxies with a one-component S\'{e}rsic $n=1$ profile and an ellipticity of $\epsilon = 0.1$ and convolved them with a Gaussian kernel adopting our average {\it WHT} $V$-band seeing point spread function (PSF) FWHM of $0.9$\,arcsec. For each model, we generated one copy of our mosaic where we inserted about $80$ duplicates, requiring that they do not overlap with each other or fall on one of the masked regions indicated in Figure~\ref{fig:fig1}. The total number of models inserted into $69$ different mosaic copies amounts to $5478$.

We ran \textsc{SExtractor} several times with varying parameter settings on the mosaics with the inserted model galaxies in order to test which parameter configuration yields the highest detection fractions. In particular, the parameters \texttt{DT} and \texttt{DMIN}, specifying the minimum number of connected pixels (\texttt{DMIN}) above a certain detection threshold (\texttt{DT}) that are required to result in a detection, control which kind of sources will be detected. 

\begin{figure*}
	\includegraphics[width=\textwidth]{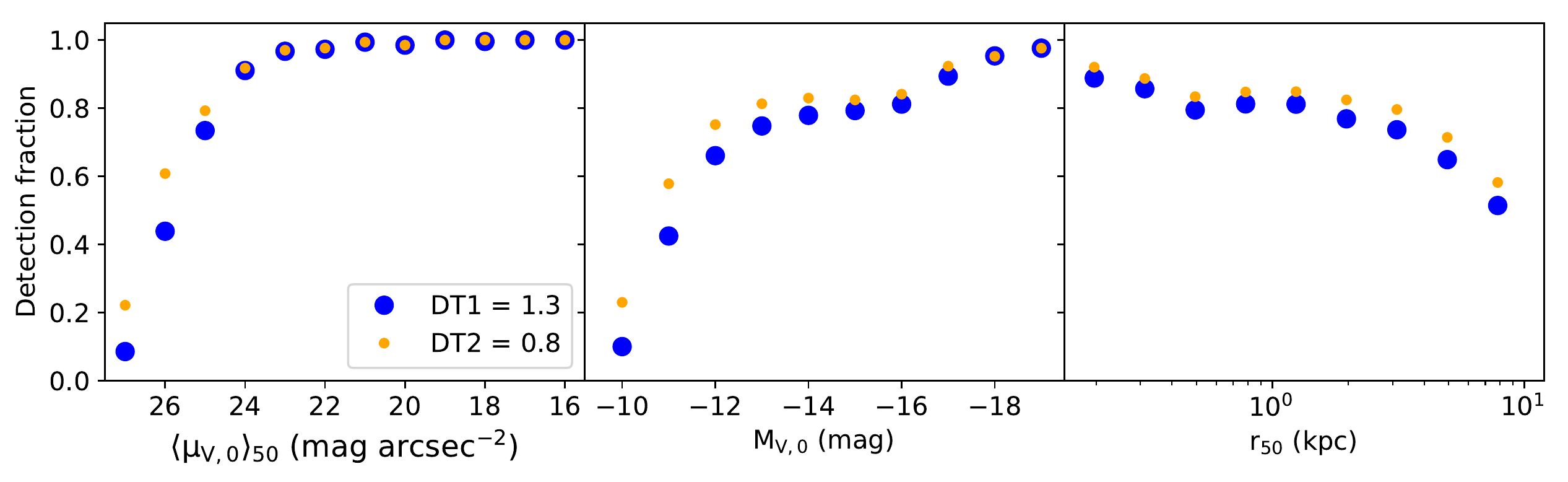}
    \caption{Comparison of the detection fractions of the inserted model galaxies for the two considered \textsc{SExtractor} configurations with detection thresholds of \texttt{DT1} $= 1.3\,\sigma$ and \texttt{DT2} $= 0.8\,\sigma$, respectively, as a function of intrinsic model parameters. All other \textsc{SExtractor} configuration parameters were adopted as given in Table~\ref{tab:tab1}. Models are considered as detected if their position matches a \textsc{SExtractor} detection within $1.5$\,arcsec.}
    \label{fig:fig2}
\end{figure*}

Figure~\ref{fig:fig2} shows that with a parameter configuration of \texttt{DT} $= 1.3\,\sigma$ and \texttt{DMIN}\,$= 25$\,pixels, we are able to detect more than $90$\,per\,cent of all model galaxies with $\langle\mu_{V,0}\rangle_{50} \leq 24$\,mag\,arcsec$^{-2}$. A lower detection threshold of \texttt{DT} $= 0.8\,\sigma$ with the same \texttt{DMIN} parameter improves the detection of fainter surface brightness sources with $\langle\mu_{V,0}\rangle_{50} > 24$\,mag\,arcsec$^{-2}$. However, although both \textsc{SExtractor} configurations yield similar detection fractions for sources with $\langle\mu_{V,0}\rangle_{50} \leq 24$\,mag\,arcsec$^{-2}$, the derived photometry of the \texttt{DT} $= 0.8\,\sigma$ detections often turns out to be less reliable. For example, sources with diffuse halos, which would be detected as single objects with the \texttt{DT} $= 1.3\,\sigma$ configuration, are frequently split up into multiple detections with the \texttt{DT} $= 0.8\,\sigma$ parameter setting. Therefore, we adopted a two-pass detection, where we first ran \textsc{SExtractor} with a detection threshold of \texttt{DT1} $= 1.3\,\sigma$, and then with a detection threshold of \texttt{DT2} $= 0.8\,\sigma$. We merged the source catalogs of both runs, where we considered all sources from the \texttt{DT1} run and additional sources from the \texttt{DT2} run that were not detected with the  \texttt{DT1} \textsc{SExtractor} parameter setting.\footnote{We considered a model galaxy as nondetected if its position did not match to any source detected by \textsc{SExtractor} within $1.5$\,arcsec.} In both cases, we ran \textsc{SExtractor} with internal filtering prior to source detection, adopting a Gaussian filter with FWHM $= 4$\,pixels, which is on the order of the average {\it WHT} $V$-band seeing PSF FWHM. To take into account the noise properties of our mosaic during the detection process, we furthermore provided a weight image generated from our data, which is described in \citet{Wittmann2017}.

The detection of faint sources close to or superimposed with brighter sources is very sensitive to the \textsc{SExtractor} deblending parameters, as well as to the subtracted background map, which is internally generated by \textsc{SExtractor}. For the deblending, we used a very low deblending contrast (\texttt{DEBLEND\_MINCONT} $=0.00001$) together with a high number of deblending thresholds (\texttt{DEBLEND\_NTHRESH} $=64$). This ensures that a bright source is split up into a sufficient number of subdetections, allowing us to recognize superimposed faint sources as separate detections. We note that this may also lead to splitting up extended low surface brightness sources. Multiple detections of a single source were, however, rejected later through visual inspection.

The properties of the subtracted \textsc{SExtractor} background map are regulated by the parameters \texttt{BACK\_SIZE}, specifying the size of the region within which the mean background is estimated, and \texttt{BACK\_FILTERSIZE}, denoting the width of the filter that is used to smooth the background map. When the size of the background box becomes comparable to the size of a certain object in the data, part of the object flux will be incorporated into the background map and will be subtracted. Thus, the \texttt{BACK\_SIZE} parameter should be small enough to subtract most of the light from extended halos of bright sources, enabling the detection of underlying faint sources, but at the same time larger than the size of the typical sources of interest. We therefore adopted a \texttt{BACK\_SIZE} parameter of $64$\,pixels, corresponding to about $15$\,arcsec or $5$\,kpc at the distance of Perseus, and a \texttt{BACK\_FILTERSIZE} of $3$.  We show the mosaic with the subtracted \textsc{SExtractor}-generated background map  that we used for source detection in Figure~\ref{fig:fig1} (bottom panel).

A summary of our adopted \textsc{SExtractor} parameters is given in Table~\ref{tab:tab1}. In total, we detected $29,111$ sources, from which $7899$ sources were only detected with the \texttt{DT1} $= 0.8\,\sigma$ \textsc{SExtractor} run. We excluded sources whose centers fall onto a masked region or are located at the edge of our mosaic with centers falling outside of the observed mosaic region.

\begin{deluxetable}{ll}
\tablecaption{Summary of the Adopted \textsc{SExtractor} Parameters for Source Detection \label{tab:tab1}}
\tablecolumns{2}
\tablewidth{0pt}
\tablehead{
\colhead{Parameter}\hspace{2.5cm} & \colhead{Value}\hspace{3cm}
}
\startdata
  \texttt{DT1} & $1.3\,\sigma$\\
  \texttt{DT2} & $0.8\,\sigma$\\
  \texttt{DMIN} & $25$\,pixels\\
  \texttt{FILTER} & Gauss, FWHM $= 4$\,pixels\\
  \texttt{DEBLEND\_MINCONT} & $0.00001$\\
  \texttt{DEBLEND\_NTHRESH} & $64$\\
  \texttt{BACK\_SIZE} & $64$\,pixels\\
  \texttt{BACK\_FILTERSIZE} & $3$\\
\enddata
\tablecomments{We built our source catalog on two \textsc{SExtractor} runs. Here \texttt{DT1} denotes the detection threshold for the first run, and \texttt{DT2} denotes the detection threshold for the second run (see Section~\ref{sec:sec3.1}).}
\end{deluxetable}

\subsection{Working Sample} \label{sec:sec3.2}

\begin{figure}
	\includegraphics[width=\columnwidth]{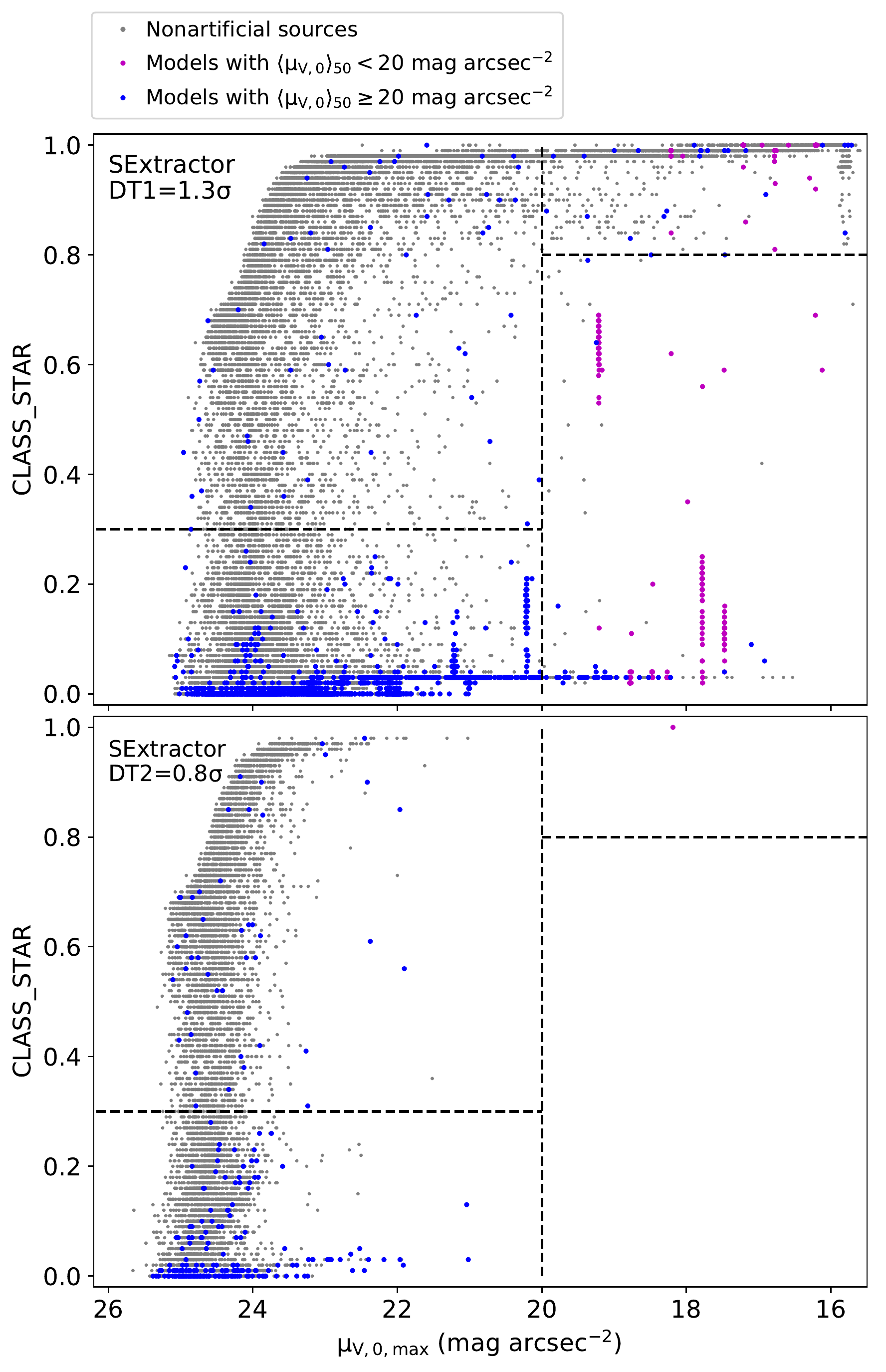}
    \caption{Separation between galaxy- and starlike sources. The figure shows the \textsc{SExtractor} parameters \texttt{CLASS\_STAR} plotted versus $\mu_{V,0,max}$ for detected model galaxies (colored dots) and detected nonartificial sources (gray dots). Low \texttt{CLASS\_STAR} values indicate extended galaxy-like sources, and high values denote compact starlike sources. Here $\mu_{V,0,max}$ denotes the maximal surface brightness measured by SExtractor, and $\langle\mu_{V,0}\rangle_{50}$ corresponds to the intrinsic surface brightness of the models, according to which they are color-coded. The top panel shows the resulting detection from the \textsc{SExtractor} run with \texttt{DT1}$= 1.3\,\sigma$, and the bottom panel shows additional detections from the \textsc{SExtractor} \texttt{DT2} $= 0.8\,\sigma$ run, which were not detected with the  \texttt{DT1} setting. The dashed lines indicate the \texttt{CLASS\_STAR} parameter cuts below which we considered the detected sources for our catalog. The cuts correspond to \texttt{CLASS\_STAR} $\leq 0.3$ for sources with $\mu_{V,0,max} \geq 20$\,mag\,arcsec$^{-2}$, and \texttt{CLASS\_STAR} $\leq 0.8$ for sources with $\mu_{V,0,max} < 20$\,mag\,arcsec$^{-2}$.}
    \label{fig:fig3}
\end{figure}

We cleaned our source catalog from unresolved starlike sources using the \textsc{SExtractor} stellarity index \texttt{CLASS\_STAR}. The index can take values between zero and one, where a value close to zero indicates an extended galaxy-like source and a value close to one indicates a compact, starlike source. Figure~\ref{fig:fig3} illustrates the \texttt{CLASS\_STAR} distribution of the inserted model galaxies and the real sources as a function of the maximal surface brightness $\mu_{V,0,max}$, measured by SExtractor. The majority ($97$ per\,cent) of our inserted model galaxies with an intrinsic $\langle\mu_{V,0}\rangle_{50} \geq 20$\,mag\,arcsec$^{-2}$ seem to be well described by \texttt{CLASS\_STAR} $\leq 0.3$. Among the more compact model galaxies with an intrinsic $\langle\mu_{V,0}\rangle_{50} < 20$\,mag\,arcsec$^{-2}$ and $r_{50} > 200$\,pc, $83$ per\,cent have \texttt{CLASS\_STAR} $\leq 0.8$. Of the compact model galaxies with $r_{50} = 200$\,pc, however, $74$\,per\,cent have \texttt{CLASS\_STAR} $> 0.8$ and are indistinguishable from unresolved point sources. Thus, for all real detected sources, we consider all sources with \textsc{SExtractor} output parameters in the regime $\mu_{V,0,max} \geq 20$\,mag\,arcsec$^{-2}$ and \texttt{CLASS\_STAR} $\leq 0.3$, as well as sources with $\mu_{V,0,max} < 20$\,mag\,arcsec$^{-2}$ and \texttt{CLASS\_STAR} $\leq 0.8$. This reduced our source catalog to a total number of $13,132$ sources, where $3980$ sources were detected by the \textsc{SExtractor} \texttt{DT2} run. 

We only included sources in our catalog with \textsc{SExtractor} flags $ \leq 3$, corresponding to unflagged sources, sources with close neighbors, and  sources that were originally blended with another object. In total, 15 sources have flags $ > 3$ and were not considered in the catalog. Since for very faint sources the photometry uncertainties and background contamination significantly increase, we furthermore excluded sources that had an extinction-corrected \textsc{SExtractor} Petrosian magnitude \citep{Petrosian1976} fainter than $M_{V,0,p} = -11$\,mag\footnote{Using a definition of $\eta = 1/0.2$ for the Petrosian radius implemented in \textsc{SExtractor}, and setting the aperture at two Petrosian radii.} at the distance of Perseus. Our final working sample comprises a total number of $7255$ sources. The parameter cuts defining the working sample are summarized in Table~\ref{tab:tab2}.

\begin{deluxetable*}{l}
\tablecaption{Definition of the Working Sample Based on \textsc{SExtractor} Parameters \label{tab:tab2}}
\tablecolumns{1}
\tablewidth{0pt}
\tablehead{
\colhead{Parameter Criterion}
}
\startdata
  ($\mu_{V,0,max} \geq 20$\,mag\,arcsec$^{-2}$ \& \texttt{CLASS\_STAR} $\leq 0.3$) $ \vert $ ($\mu_{V,0,max} < 20$\,mag\,arcsec$^{-2}$ \& \texttt{CLASS\_STAR} $\leq 0.8$)\\
  \textsc{SExtractor} flag $ \leq 3$\\
  $M_{V,0,p} \leq -11$\,mag\\
\enddata
\end{deluxetable*}

\subsection{Completeness} \label{sec:sec3.3}

\begin{figure*}
	\includegraphics[width=\textwidth]{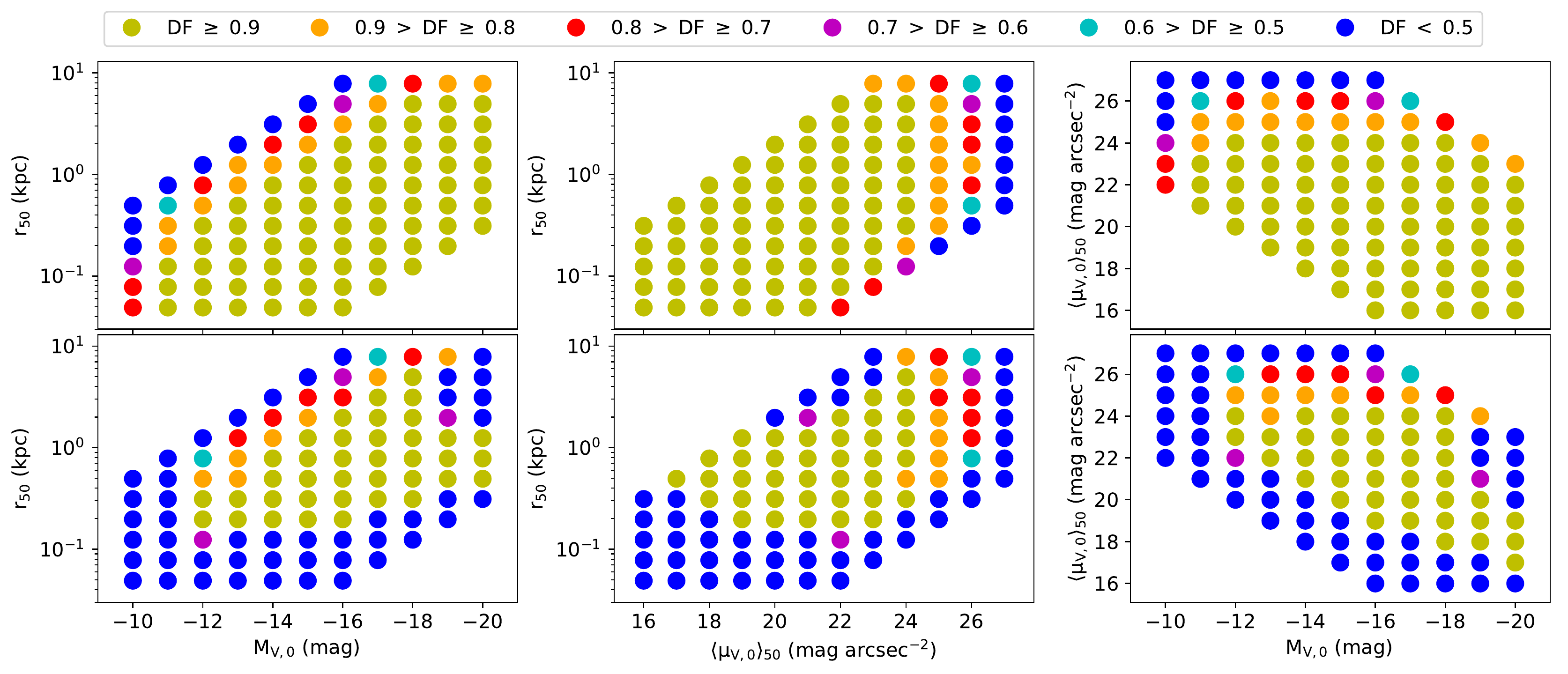}
    \caption{Completeness estimate based on the detection of model galaxies in the {\it WHT} $V$-band mosaic. The plots show the detection fractions (DF) of detected-to-inserted model galaxies as a function of model input parameters. In total, $8015$ models were inserted into $101$ copies of the mosaic. The first row displays the completeness achieved with the \textsc{SExtractor} source detection. The second row illustrates the completeness after having applied the rejection criteria to define our working sample (see Section~\ref{sec:sec3.1} and \ref{sec:sec3.2}).}
    \label{fig:fig4}
\end{figure*}

Figure~\ref{fig:fig4} displays a completeness estimate of our catalog based on the inserted model galaxies. Shown are the detection fractions, which are the ratios of the number of detected models to the total number of inserted models, as a function of the model input parameters. For the completeness estimate, we extended the parameter range of the inserted models to brighter magnitudes of $M_{V,0} = -20$\,mag and smaller sizes of $r_{50} = 50$\,pc in order to cover the full observed parameter range of our final catalog. In total, a number of 101 models with different parameters contribute to the completeness estimate, where each model type was inserted several times into one copy of the mosaic, amounting to a total number of  $8015$ inserted models.

In the first row of Figure~\ref{fig:fig4}, we show the detection fractions achieved with the combined \textsc{SExtractor} runs with \texttt{DT1} $= 1.3\,\sigma$ and \texttt{DT2} $= 0.8\,\sigma$. It can be seen that the completeness drops below $50$\,per\,cent for models with $\langle\mu_{V,0}\rangle_{50} = 27$\,mag\,arcsec$^{-2}$ and $M_{V,0} = -10$\,mag. The second row displays the resulting detection fractions after having applied the source rejection criteria to define our working sample (see Section~\ref{sec:sec3.1} and \ref{sec:sec3.2}). Removing starlike sources by applying the \texttt{CLASS\_STAR} parameter cut results in detection fractions lower than $50$\,per\,cent for the models with the smallest sizes ($r_{50} < 150$\,pc) and the brightest surface brightnesses ($\langle\mu_{V,0}\rangle_{50} \leq 17$\,mag\,arcsec$^{-2}$). Excluding sources with $M_{V,0,p} > -11$\,mag measured by \textsc{SExtractor} implies detection fractions lower than $50$\,per\,cent for models with intrinsic $M_{V,0} = -11$\,mag, since \textsc{SExtractor} tends to underestimate the magnitudes of the detected models. Rejecting sources that would get masked, and therefore excluded, with the bright object mask displayed in Figure~\ref{fig:fig1} affects the brightest and largest sources with $M_{V,0} \leq -19$\,mag and $r_{50} > 2$\,kpc, where the average detection fraction is lower than $50$\,per\,cent. In summary, the average completeness of our catalog is 92\,per\,cent in the parameter range $-12 \geq M_{V,0} \geq -18$\,mag, $18 \leq \langle\mu_{V,0}\rangle_{50} \leq 26$\,mag\,arcsec$^{-2}$, and $0.2 \leq r_{50} \leq 7.8$\,kpc.

\section{Photometry} \label{sec:sec4}

Our aim was to use \textsc{galfit} \citep{Peng2002, Peng2010} for measuring the photometry and structural parameters of our working sample in the {\it WHT} $V$-band mosaic. We first derived Petrosian magnitudes, half-light radii, and concentration values, which we used as first-guess parameters for \textsc{galfit}, since the photometry provided by \textsc{SExtractor} is often inaccurate for extended sources. We visually examined all sources to check the performance of our photometric measurements,\footnote{Done by CW.}, thereby rejecting too-heavily contaminated sources and false detections.

We performed the photometric measurements on the partly object-subtracted {\it WHT} mosaic (see Figure\ref{fig:fig1}, top right panel) and cut out a postage stamp image for each source with a size of $701 \times 701$\,pixels. We then generated masks of neighboring sources using \textsc{SExtractor}. We ran \textsc{SExtractor} on the entire mosaic, adopting the same parameter settings as specified in Table~\ref{tab:tab1} using \texttt{DT1}, but with a larger \texttt{BACK\_SIZE} parameter of $256$\,pixels, since this results in larger masked areas for extended sources. For each source, we generated a postage stamp mask, where we unmasked the respective source.

\subsection{Petrosian Photometry} \label{sec:sec4.1}

For the photometric measurements, we defined the Petrosian radius $R_p$ such that the Petrosian index $\eta(R_p) = \langle I \rangle_{R_p} / I(R_p) = 1/0.3$, where $\langle I \rangle_{R_p}$ denotes the average intensity within $R_p$, and $I(R_p)$ gives the isophotal intensity at $R_p$ \citep{Graham2005b}. We adopted a value of $\eta(R_p) = 1/0.3$ instead of the more commonly used $\eta(R_p) = 1/0.2$, since the former proved to be more robust for low surface brightness sources to prevent contamination through close neighbor objects or uneven background. We measured the total flux within a circular aperture of $1.5\,R_p$ and derived the corresponding half-light radius from it. We estimated S\'{e}rsic indices from the measured concentration values $r_{90}/r_{50}$, where we matched the observed concentration values to the values calculated for analytic S\'{e}rsic profiles with $n=0.5 - 4.0$. For observed concentration values lower or higher than the calculated ones, we adopted a S\'{e}rsic index of $n=0.5$ or $4$, respectively.

As a rough estimate for the local background underlying a source, we used \textsc{SExtractor}-generated background maps. For sources without close bright and extended neighbor sources, we applied a \textsc{SExtractor} background map with a large \texttt{BACK\_SIZE} parameter of $256$\,pixels, corresponding to $60.7$\,arcsec or $20.6$\,kpc. For small sources in the vicinity of bright extended sources, we used a background map with a smaller \texttt{BACK\_SIZE} parameter of $32$\,pixels, corresponding to $7.6$\,arcsec or $2.6$\,kpc, in order to remove the background gradient of the neighboring source.\footnote{In total, we applied the \texttt{BACK\_SIZE} $=32$\,pixel background subtraction to 1962 sources. These all have very small sizes with Petrosian half-light radii $r_{50,p} \leq 3.6$\,arcsec ($\widehat{=} 1.34$\,kpc) and with an average value of $r_{50,p} = 0.85$\,arcsec ($\widehat{=} 290$\,pc).} In the case of larger sources with contaminating neighbors, we fitted and subtracted the light distribution of the respective neighbor source with \textsc{iraf} \textit{ellipse} in order to avoid the \textsc{SExtractor} background map, which may subtract part of the source flux.

We carefully inspected the measured Petrosian apertures visually for each of the $7255$ sources from our working sample in order to identify cases in which the aperture is obviously too large with regard to the visible extent of the source. Most often, problems arose due to imperfect masking, when either the automated unmasking of the source failed, such that large parts of the source were still masked, or when neighboring sources contaminated the flux measurement, since they were insufficiently masked. We therefore manually adjusted the masks of all affected sources and measured the Petrosian parameters again.

In the following cases, we entirely excluded a source from any further analysis: 
\begin{enumerate}
\item The source is heavily blended into another source where neither a larger mask nor a \textsc{SExtracor} background subtraction with \texttt{BACK\_SIZE} $= 32$\,pixels would result in reliable photometric measurements. We note that the majority of sources in this category appeared very small in size, being possible interlopers suffering from wrong \textsc{SExtractor} photometry and / or \texttt{CLASS\_STAR} measurements. In the case of extended galaxy-like sources, we tried our best to fit and subtract the contaminating neighboring source.
\item The source is barely visible in our data, although it has been given a \textsc{SExtractor} magnitude of $M_{V,0,p} \leq -11$\,mag.
\item The source forms part of another source, i.e., the stripped material or spiral arms of luminous cluster galaxies, that is not a subject of this study.
\item The source is likely an artifact resulting from reflections or stray light in the data.
\item In the case of multiple detections of a single source, we derived Petrosian photometry for the detection that visually appeared best centered on the source and rejected the remaining duplicate detections.
\end{enumerate}

In total, we successfully derived Petrosian photometry for $6038$ sources and excluded $1217$ sources from the further analysis.

\subsection{WHT $V$-band PSF} \label{sec:sec4.2}

\begin{figure*}
	\includegraphics[width=\textwidth]{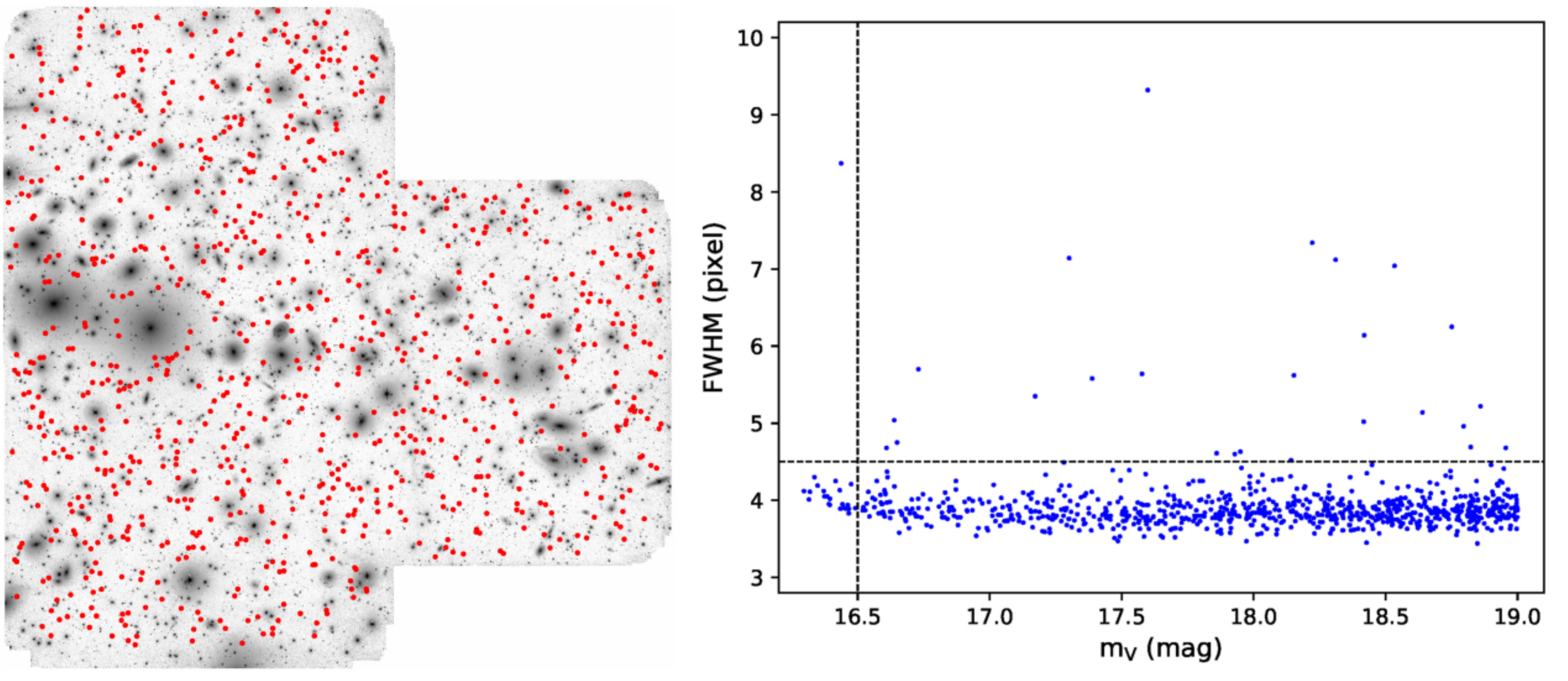}
    \caption{{\it Left:} distribution of PSF star candidates (red dots) in the {\it WHT} $V$-band mosaic. {\it Right:} FWHM measured by \textsc{SExtractor} versus apparent magnitude measured by \textsc{iraf} \textit{phot}. For the average stacked PSF used by \textsc{galfit}, we considered all stars with FWHM $< 4.5$\,pixels and $m_{V} > 16.5$\,mag, as indicated by the dashed lines.}
    \label{fig:fig5}
\end{figure*}

In order to run \textsc{galfit} with PSF deconvolution, we first generated a PSF from our data using routines from the \textsc{iraf} package \textsc{daophot}. We ran \textsc{SExtractor} tuned to detect small and compact sources\footnote{With the configuration \texttt{DT} $= 2.5\,\sigma$, \texttt{DMIN} $= 5$\,pixels, \texttt{DEBLEND\_NTHRESH} $= 32$, \texttt{DEBLEND\_MINCONT} $= 0.01$, \texttt{BACK\_SIZE} $= 64$\,pixels, and \texttt{BACKFILTER\_SIZE} $= 3$.} and selected PSF star candidates by requiring that the sources have \texttt{CLASS\_STAR} $> 0.9$ and not be flagged. We furthermore rejected sources that are superimposed on bright extended sources by excluding all PSF star candidates that overlap with a mask generated with \textsc{SExtractor}\footnote{Using the configuration \texttt{DT} $= 5\,\sigma$, \texttt{DMIN} = $10,000$\,pixels, \texttt{DEBLEND\_NTHRESH} = $32$, \texttt{DEBLEND\_MINCONT} = $0.01$, \texttt{BACK\_SIZE} = $256$\,pixels, and \texttt{BACKFILTER\_SIZE} = $3$.}. This mask is much larger than the one shown in Figure~\ref{fig:fig1} in order to yield the cleanest possible sample of PSF star candidates. We also excluded PSF star candidates that fall on very low S/N regions in our mosaic, where the pixels have a weight lower than $200$, corresponding to $\sigma \sim 0.07$\,ADU. We subsequently ran the \textsc{iraf} task \textit{phot} to perform aperture photometry\footnote{Using an aperture with a radius of $r = 5$\,pixels ($\mathrel{\widehat{=}} 1.2$\,arcsec) and a background annulus with a width of $5$\,pixels at $r = 15$\,pixels.} and selected all stars with apparent $V$-band magnitudes brighter than $m_{V} = 19$\,mag that were not saturated. This resulted in a selection of $845$ PSF stars in our mosaic. In Figure~\ref{fig:fig5}, we display their spatial and their FWHM distribution. The FWHM distribution indicates a few outliers among the selected stars with large FWHM values. A visual examination revealed that most of them are partly unresolved double sources. We therefore rejected sources with  FWHM $> 4.5$\,pixels. In addition, Figure~\ref{fig:fig5} shows a slight upward trend in FWHM toward brighter magnitudes. As a consequence we only considered stars with $m_{V} > 16.5$\,mag. The final sample of PSF stars comprises $797$ stars with an average FWHM of $3.86$\,pixels, as measured by \textsc{SExtractor}. We used this PSF star sample to construct a model PSF with the \textsc{iraf} task \textit{psf}, which consists of an analytic component obtained by a fit to all of the stars in the sample plus a lookup table quantifying the deviations of the analytic function from the empirical PSF.

\subsection{\textsc{galfit} Photometry} \label{sec:sec4.3}

\begin{deluxetable*}{lll}
\tablecaption{Photometry Processing Overview of Our Working Sample \label{tab:tab3}}
\tablecolumns{2}
\tablewidth{0pt}
\tablehead{
\colhead{Photometry Processing} & \colhead{Number of Sources} & \colhead{pflag}
}
\startdata
  \textsc{galfit}: S\'{e}rsic fit with free parameters & $3161$ & 1\\
  \textsc{galfit}: S\'{e}rsic fit with free parameters but constrained $n=4$& $80$ & 2\\
  \textsc{galfit}: S\'{e}rsic fit with fixed $n$ & $1238$ & 3\\
  \textsc{galfit}: S\'{e}rsic fit with fixed $n$ and position & $19$ & 4\\
  \textsc{galfit}: S\'{e}rsic fit with fixed $n$ and $b/a$ & $92$ & 5\\
  \textsc{galfit}: S\'{e}rsic fit with fixed $n$, $r_{50}$, and $b/a$ & $332$ & 6\\
  \textsc{galfit}: S\'{e}rsic fit with fixed $n$, $b/a$, and position& $6$ & 7\\
  \textsc{galfit}: S\'{e}rsic fit with fixed $n$, $r_{50}$, $b/a$, and position & $17$ & 8\\
  Petrosian $M_v, r_{50}, n$ & $492$ & 9\\
  No photometry\tablenotemark{a} & $1818$ & \nodata\\
\enddata
\tablecomments{Here ``pflag'' denotes the photometry flag given in the final catalog.}
\tablenotetext{a}{We do not provide photometric measurements for sources excluded according to the criteria specified in Section~\ref{sec:sec4.1}, sources with $M_{V,0} > -10$\,mag, or sources that were excluded or revealed as double sources in the {\it Subaru} imaging data (see Section~\ref{sec:sec5}).}
\end{deluxetable*}

We derived structure parameters with \textsc{galfit} for the $6038$ nonexcluded sources from our working sample. We ran \textsc{galfit} for each source with PSF deconvolution, object masks, and a $\sigma$ image generated from the corresponding weight image of the {\it WHT} mosaic, where $\sigma = 1 / \sqrt{\mathrm{weight}}$. We fitted each source with a one-component S\'{e}rsic profile and used the derived Petrosian magnitude, half-light radius, and estimated  S\'{e}rsic index, as well as the \textsc{SExtractor} source position, axis ratio and position angle, as first-guess parameters for \textsc{galfit}. We simultaneously fitted the sky component, which can account for both a background offset and a gradient. For sources where we previously subtracted the \textsc{SExtractor}-generated background map with a \texttt{BACK\_SIZE} of $32$\,pixels, we adopted this background instead of fitting the background with \textsc{galfit}, since only the former eliminates small-scale background variations introduced by contaminating neighboring sources (see Section~\ref{sec:sec3.1}). For $25$ sources, we used the previously subtracted \textsc{SExtractor} \texttt{BACK\_SIZE} $ = 256$\,pixel background map for \textsc{galfit}, since simultaneously fitting the source and the background did not succeed. We specify the applied background subtraction method in the final catalog.

In the first iteration with \textsc{galfit}, we performed the S\'{e}rsic fit with seven free parameters\footnote{This included the $x$-, $y$-position, magnitude, half-light radius, S\'{e}rsic index, axis ratio and position angle.}, but constrained the S\'{e}rsic index to $n \leq 4$. We note that due to the limited resolution of our data, we are not capable of reliably discriminating between a source with a S\'{e}rsic index around $n = 4$ and a larger value, due to the small change in the profile shape for large S\'{e}rsic $n$ \citep[see, e.g.,][Figure\,1]{Graham2005a}. We visually examined all fitted models by comparing them to the respective source, as well as the residual images. Some sources showed significant residuals after the subtraction of the model, although the fitted model seemed to provide a good description of the overall shape of the source. We adopted the photometric parameters but flagged the respective sources in the final catalog. For about one-third of the sample, the S\'{e}rsic model with free parameters did not provide a good fit, or the fit did not converge.

In these cases, we performed a second iteration with \textsc{galfit} where we held the S\'{e}rsic index fixed at the value estimated from the concentration $r_{90}/r_{50}$ (see Section~\ref{sec:sec4.1}) to stabilize the fit. Subsequent visual inspection yielded good solutions for about $1200$ of the sources. We also compared the models of all sources that were previously fitted with free parameters but constrained $n=4$ and adopted the solution, which yielded less pronounced residuals.

We refitted all sources where our fitting attempts had not succeeded in a third iteration with \textsc{galfit}. In this iteration, we held the $x$-, $y$-position fixed, in addition to the S\'{e}rsic index. This succeeded for another $19$ sources. For the remaining sources, the \textsc{galfit} solutions did not converge, since the half-light radius and / or the axis ratio fell below the lowest parameters of $r_{50} = 0.5$\,pixels and $b/a = 0.1$ accepted by \textsc{galfit} to fit a source. We therefore ran \textsc{galfit} in a fourth iteration by adopting the fixed lowest values for $r_{50}$ and / or $b/a$ in addition to the fixed S\'{e}rsic index. For sources that were fitted with a fixed $r_{50} = 0.5$\,pixels, we also adopted a fixed axis ratio of $b/a=1$ due to the high uncertainties of the latter at these small sizes.

For $270$ sources, it was not possible to obtain a good fit with \textsc{galfit}, but the derived Petrosian parameters seemed to provide a good estimate of the structural parameters. In most of these cases, the affected sources were very faint or showed strong residuals due to a very complex structure where a single S\'{e}rsic fit might not be a good approximation. We also adopted Petrosian photometry for $222$ sources fitted with \textsc{galfit} but that have statistical uncertainties in $m_{V,0}$ and / or $\langle\mu_{V,0}\rangle_{50}$ larger than $1$\,mag. These are almost exclusively very faint sources, with $90$\,per\,cent of them having $M_{V,0} > -12$\,mag. We note that the half-light radii derived from Petrosian photometry only provide upper size limits for small sources, since the measurements did not involve a PSF deconvolution. We entirely excluded sources with absolute $V$-band magnitudes fainter than $M_{V,0} = -10$\,mag from our catalog. For the position angle measurements, we only list values for sources with $b/a < 0.9$, due to the high uncertainties for nearly round sources. We provide an overview of the photometry processing of our working sample in Table~\ref{tab:tab3} and include the photometric measurements in the final catalog.

We calculated the effective surface brightness $\langle\mu_{V}\rangle_{50}$ within the half-light radius from the measured parameters and obtained the Galactic foreground extinction $A_V$ at the position of each source from the IRSA Galactic Reddening and Extinction Calculator,\footnote{We acknowledge the use of the NASA/ IPAC Infrared Science Archive, which is operated by the Jet Propulsion Laboratory, California Institute of Technology, under contract with the National Aeronautics and Space Administration.}, with reddening maps from \citet{Schlafly2011}. 

We provide the statistical uncertainties estimated by \textsc{galfit} based on the pixel noise in the final catalog. The uncertainties in $\langle\mu_{V}\rangle_{50}$ were calculated from error propagation, accounting for uncertainties in $m_{V,0}$ and $r_{50}$. Parameters that were held fixed during a fit have no error estimates, as well as sources where we adopted the Petrosian photometry. We provide an estimate of the typical uncertainties of the Petrosian photometry in Figure~\ref{fig:figA1} in Appendix~\ref{sec:secA1}.

\subsection{Aperture Colors} \label{sec:sec4.4}

We derived ($g-r$) and ($r-z$) aperture colors from the {\it Subaru} data set, which we resampled to match the {\it WHT} mosaic pixel scale of 0.237\,arcsec. For all sources, we used our previously generated object masks from the WHT data but extended them,\footnote{By smoothing the mask images, which contain only values of zero and one, with a Gaussian kernel with $\sigma=1$\,pixel and then generating new masks based on the smoothed ones where all pixels with values larger than 0.01 are masked.}, since the {\it Subaru} data are deeper than the {\it WHT} data and object flux becomes visible beyond the masked areas.

We measured the colors as flux ratios within apertures with a radius of $r_{aper} = r_{50}$ and a shape defined by the previously derived ellipticity and position angle. For very small sources with  $r_{50} < 4$\,pixels, we adopted circular apertures with  $r_{aper} = 4$\,pixels, since for smaller apertures, seeing effects start to influence the color values. For sources with $r_{50} \geq 4$\,pixels, we additionally measured an "outer" color, defined as a flux ratio within an aperture with a radius of $r_{aper} = 2\,r_{50}$.

Prior to measuring the colors, we determined and subtracted the local background around each source in each of the bands. For the sources with fixed apertures of $r_{aper} = 4$\,pixels, we calculated the median background level within a circular annulus with an inner radius of $12$\,pixels and a width of $4$\,pixels. For sources with apertures scaling with their half-light radius, we adopted a background annulus following the shape of the color aperture and with a width of one $r_{50}$. We set the inner radius of the background annulus depending on the surface brightness of the respective source. Adopting the same background annulus for all sources would result in a background annulus that covers a significant amount of object flux for the high surface brightness sources due to their smaller half-light radii, whereas for the low surface brightness sources, it would be located too far from the source center due to their larger half-light radii. Therefore, we set the inner radius of the background annulus for low surface brightness sources with $\langle\mu_{V,0}\rangle_{50} \geq 24$\,mag\,arcsec$^{-2}$ at $3\,r_{50}$, for sources with $22 \leq \langle\mu_{V,0}\rangle_{50} < 24$\,mag\,arcsec$^{-2}$ at $4\,r_{50}$, and for high surface brightness sources with $\langle\mu_{V,0}\rangle_{50} \leq 22$\,mag\,arcsec$^{-2}$ at $5\,r_{50}$.

We calculated statistical uncertainties for the ($g-r$) and ($r-z$) colors based on standard error propagation as 
\begin{equation}
\Delta (g-r) = \sqrt{\left( \frac{2.5}{I_g ln10} \Delta I_g \right)^2 + \left( \frac{2.5}{I_r ln10} \Delta I_r \right)^2 }
\end{equation}
where $I_r$ and $I_g$ are the aperture flux in the $r$- and $g$-band, respectively, and $\Delta I_r$ and $\Delta I_g$ are estimated from the {\it Subaru} $r$- and $g$-band variance image,\footnote{where the same regions as in the object image are masked}, accounting for the statistical uncertainty in the object and background flux. The same applies for $\Delta (r-z)$. We only included sources with $\Delta (g-r) \leq 1$\,mag and $\Delta (r-z) \leq 1$\,mag and without saturated pixels within 3\,arcsec from the respective source center in our catalog. We included but flagged all sources that contain pixels with bleeding trails of bright stars within 3\,arcsec and may have contaminated colors.

\begin{figure}
    \centering
    \includegraphics[width=\columnwidth]{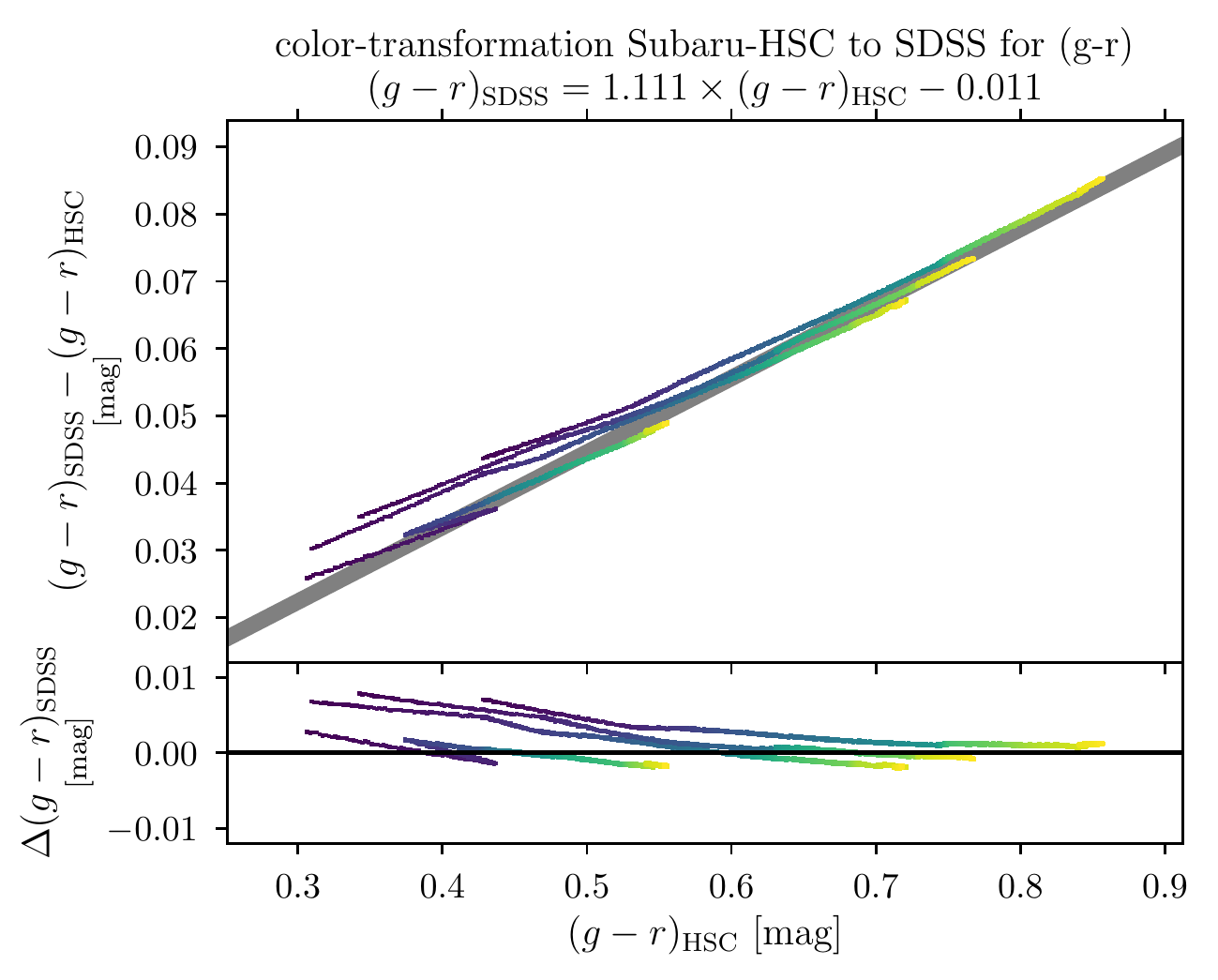}
    \includegraphics[width=\columnwidth]{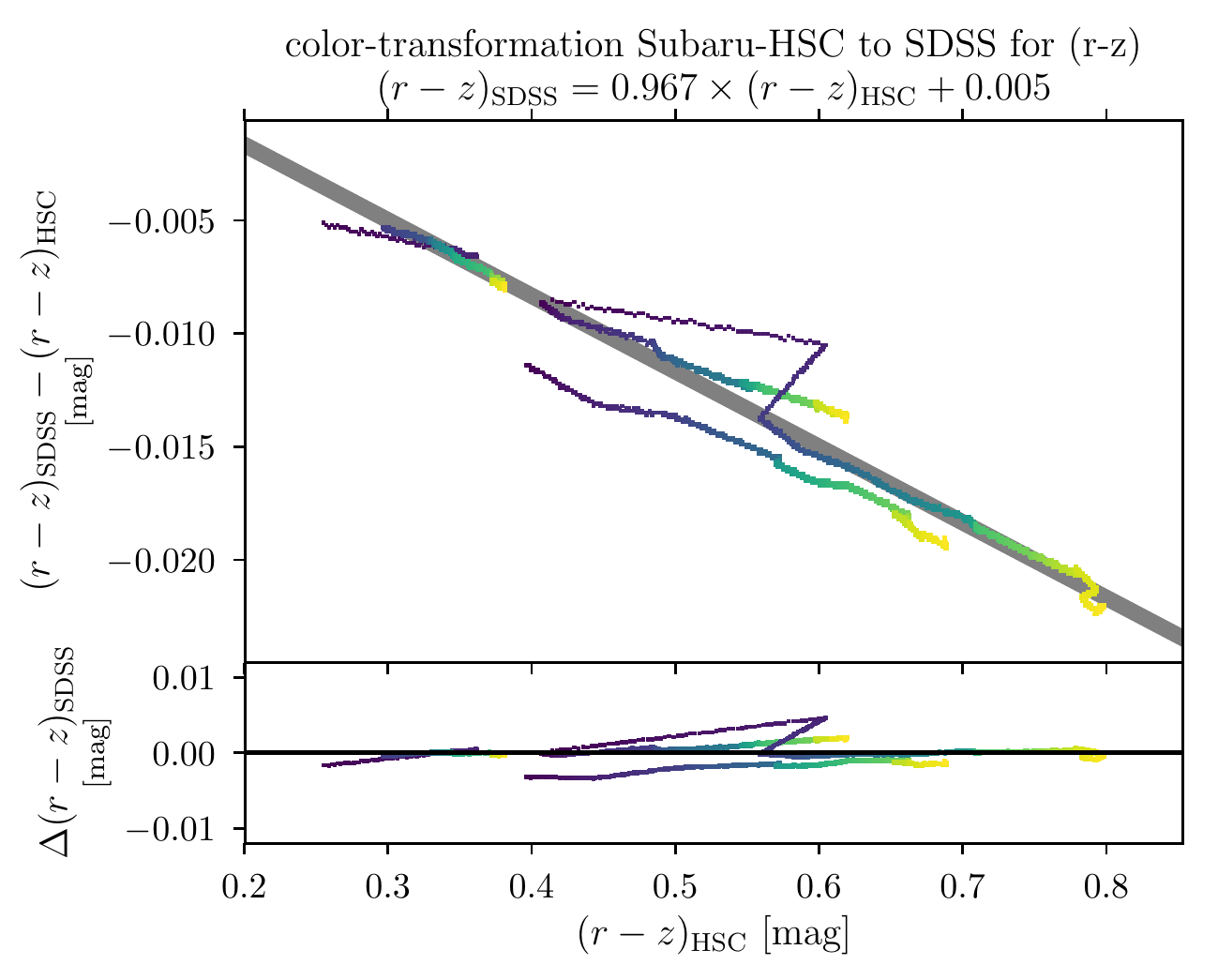}
    \caption{Color transformations from the native {\it Subaru} HSC system into the standard SDSS system for $(g-r)$ (top panel) and $(r-z)$ (bottom panel). The top part of each plot shows the difference in colors as a function of the native HSC color, while the smaller bottom part shows the residuals after accounting for the best-fit linear color transformation.}
    \label{fig:colortransformations}
\end{figure}

To enable comparison of our measured galaxy colors with data from the literature, we transformed all observed $(g-r)$ and $(r-z)$ colors from the native {\it Subaru HSC} system to the standard SDSS system. Color terms were derived synthetically using a set of GALEV stellar population models \citep{Kotulla2009} as template spectra. These were convolved with both the SDSS filters as presented by \citet{Doi2010} and the HSC filter curves. Based on the derived colors, we computed the following color transformation equations (also see Figure~\ref{fig:colortransformations} for color comparisons and residuals):

\begin{equation}
    (g-r)_{SDSS} = 1.111 \times (g-r)_{HSC} - 0.011
\end{equation}
    
\begin{equation}
    (r-z)_{SDSS} = 0.967 \times (r-z)_{HSC} + 0.005
\end{equation}

We performed an extinction correction for all three {\it Subaru} bands using the NED extinction calculator, which is based on the reddening maps of \citet{Schlafly2011} that we also used for the extinction correction in the $V$-band data.

\section{Morphological analysis} \label{sec:sec5}

\begin{deluxetable*}{ll}
\tablecaption{Morphological Classification of Our Working Sample \label{tab:tab4}}
\tablecolumns{2}
\tablewidth{0pt}
\tablehead{
\colhead{Morphological Category} & \colhead{Number of Sources}}
\startdata
(1) dE/ETG cluster candidate & $496$\\
\phantom{1) }$-10 > M_{V,0} \geq -18$\,mag & $478$\\
\phantom{1) }$-18 > M_{V,0} \geq -20.2$\,mag & $18$\\
(2) Likely background ETG or source with unresolved substructure & $3008$\\
(3) Cluster or background galaxy with late-type morphology & $384$\\
(4) Cluster or background galaxy with possibly weak substructure & $477$\\
(5) Likely cluster or background edge-on disk galaxy & $1049$\\
(6) Likely merging system in the background& $23$\\
(7) Excluded source & $283$\\
\enddata
\end{deluxetable*}

As already realized by \citet{Binggeli1985}, the early-type dwarf galaxy population of a galaxy cluster can be relatively well separated from  background ETGs or LTGs due to their distinct morphological properties. Dwarf ellipticals (dEs) differ from giant ellipticals in the way that their light distribution is more diffuse, with a generally lower S\'{e}rsic index. Thus, even though a background giant and a cluster dE galaxy may be indistinguishable in terms of their apparent luminosity and smooth morphology, the dwarf galaxy will stand out due to its more diffuse  appearance. In contrast, LTGs can appear similarly diffuse as dEs, being as well characterized by a lower S\'{e}rsic index. However, in this case, the dEs can be discriminated due to their smooth and featureless morphology.

In the following, we present a morphological analysis of our working sample based on our auxiliary {\it Subaru} data set. We used the {\it Subaru} $r$-band data with the pixel scale of 0.168\,arcsec\,pixel$^{-1}$ due to the excellent image quality with a $0.5$\,arcsec seeing PSF FWHM, allowing us to identify substructures even in relatively small sources. In addition, we generated ($g-r$)$_0$ color maps for each source in order to better separate between background ellipticals and more compact cluster dEs that are harder to discriminate based on morphology alone but clearly differ in color, with the background galaxies being redder.\footnote{We note that compact elliptical galaxies can have very red colors, too, and likely will be classified as background galaxies by our approach (also see Section~\ref{sec:sec7.1}).} The color maps also helped to distinguish between LTGs and early-type cluster candidates in the small size and faint luminosity regime, where morphological substructure in LTGs may be unrecognized due to resolution limits. While the majority of cluster dEs are commonly characterized by shallow color gradients of $g-r \lesssim \pm 0.1$\,mag \citep{Urich2017}, LTGs often display stronger gradients, indicative of the presence of young stars within the galaxy \citep{Gonzalez-Perez2011}.

We established the following categories according to the morphological and color properties of our working sample:

\begin{enumerate}
\item dE/ETG cluster candidate (496 sources)\\
The source is characterized by a smooth morphology and a diffuse optical appearance. It has a regular, symmetric shape and does not show substructure like spiral arms or clumps. The source may, however, contain a disk, bulge, or nucleus. The ($g-r$)$_0$ color map indicates a color of $(g-r)_0 \lesssim 1$, typical for early-type cluster galaxies \citep[cf.][]{Janz2009}, without obvious color substructure. We furthermore distinguished between likely ($398$) and possible ($98$) dE/ETG cluster candidates, where the latter includes less convincing candidates based on either morphological or color properties. We note that $18$ candidates have $M_{V,0} < -18$\,mag. We did not, however, want to generate separate classes for giant and dwarf ETGs, since it is difficult to separate galaxies brighter and fainter than a given luminosity limit from each other based on a visual classification.

\item Likely background ETG or source with unresolved substructure (3008 sources)\\
The source has an early-type morphological appearance but is either characterized by a very red color, indicative of being a massive elliptical galaxy in the background, or shows a color substructure, such as a very red center with bluer outskirts. The latter might point to a face-on disk galaxy with a star forming bluer disk where the substructure is unresolved. Also, sources with very small sizes fall into this category, where insufficient resolution limits a morphological analysis. 

\item Cluster or background galaxy with late-type morphology (384 sources)\\
The source shows clear LTG features like spiral arms or clumpy morphology. Some sources have a very asymmetric or irregular shape. 

\item Cluster or background galaxy with possibly weak substructure (477 sources)\\
The source does not unambiguously have enough substructure to be part of category 3, but it is also not smooth enough to be part of class 2. Either the source may contain only weak substructure or the substructure may not be revealed as clearly as in category 3 due to the small size of the source and the limited resolution in the data. 

\item Likely cluster or background edge-on disk galaxy (1049 sources)\\
The source is characterized by a very elongated shape, indicative of being an edge-on disk galaxy. The color map of these sources often shows a characteristic substructure with a red center and bluer outskirts, possibly pointing to a late-type star-forming disk galaxy. This category might, however, also contain early-type disk galaxies, since the edge-on view often does not allow one to robustly distinguish between the two morphological classes. There might be some overlap with category 2, with some sources possibly being very elongated background elliptical galaxies. 

\item Likely merging system in the background (23 sources)\\
The source seems to be merging or interacting with another source.

\item Excluded source (283 sources)\\
The source is identified as a double or multiple source in the higher-resolution {\it Subaru} data ($264$ sources) or revealed as an image artifact in the {\it WHT} data, as confusion with Galactic cirrus, or is part of the star-forming filaments of the NGC\,1275 system ($19$ sources). The  double or multiple sources could be either physically close sources, which are possibly merging at high redshift, or projections. We excluded all sources of this class from our final catalog.

\end{enumerate}

\begin{figure*}
	\includegraphics[width=\textwidth]{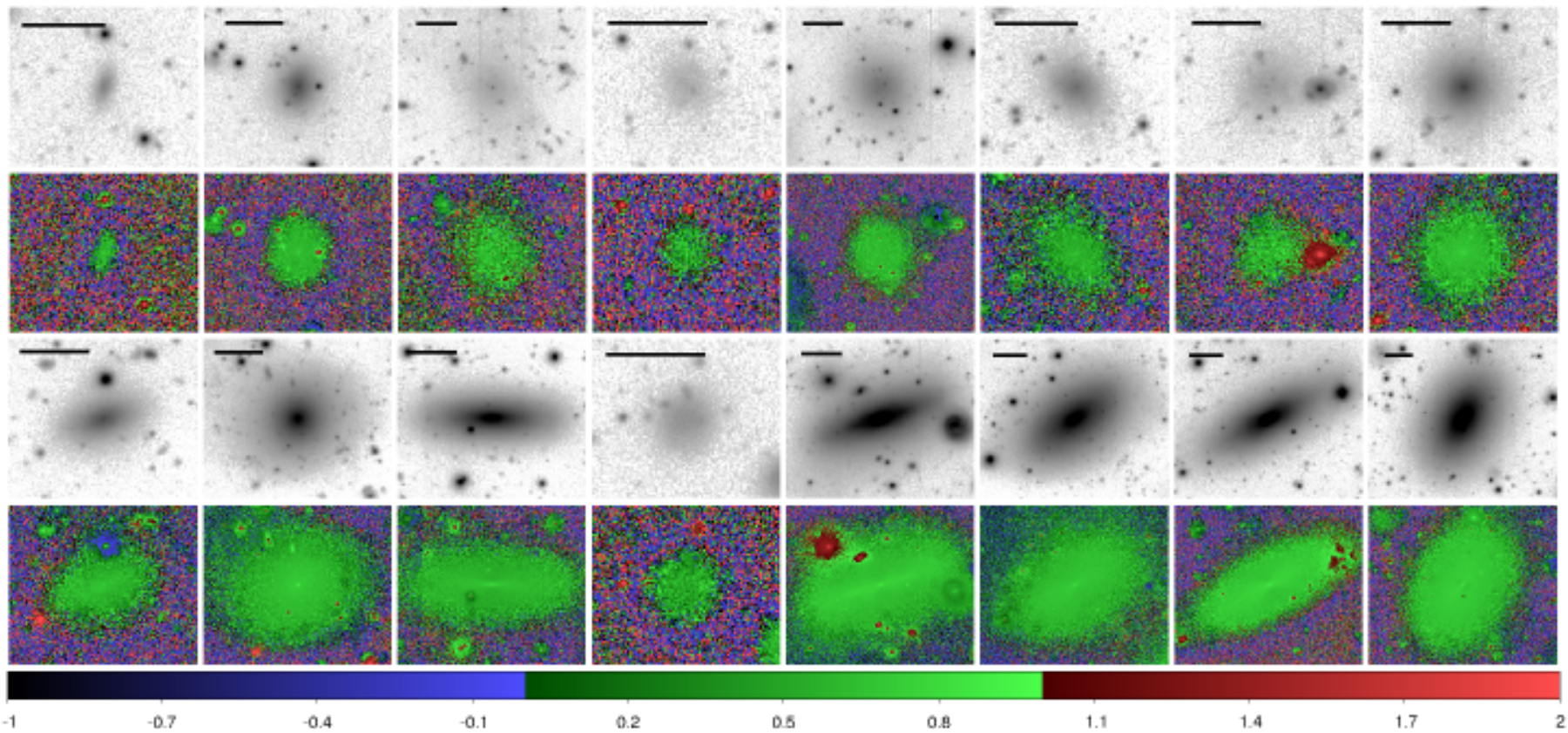}
    \caption{Perseus cluster dE/ETG candidates. For each candidate, we show the {\it Subaru} $r$-band image in the top panel and the $(g-r)_0$ color map in the bottom panel. The scale bar in each $r$-band image corresponds to a length of $10$\,arcsec. The color bar at the bottom of the figure indicates the $(g-r)_0$ values shown in the color maps. The candidates displayed in the fifth through eighth panels in the bottom row have $M_{V,0} < -18$\,mag, whereas the other candidates in the figure are fainter.}
    \label{fig:fig6a}
\end{figure*}

\begin{figure*}
	\includegraphics[width=\textwidth]{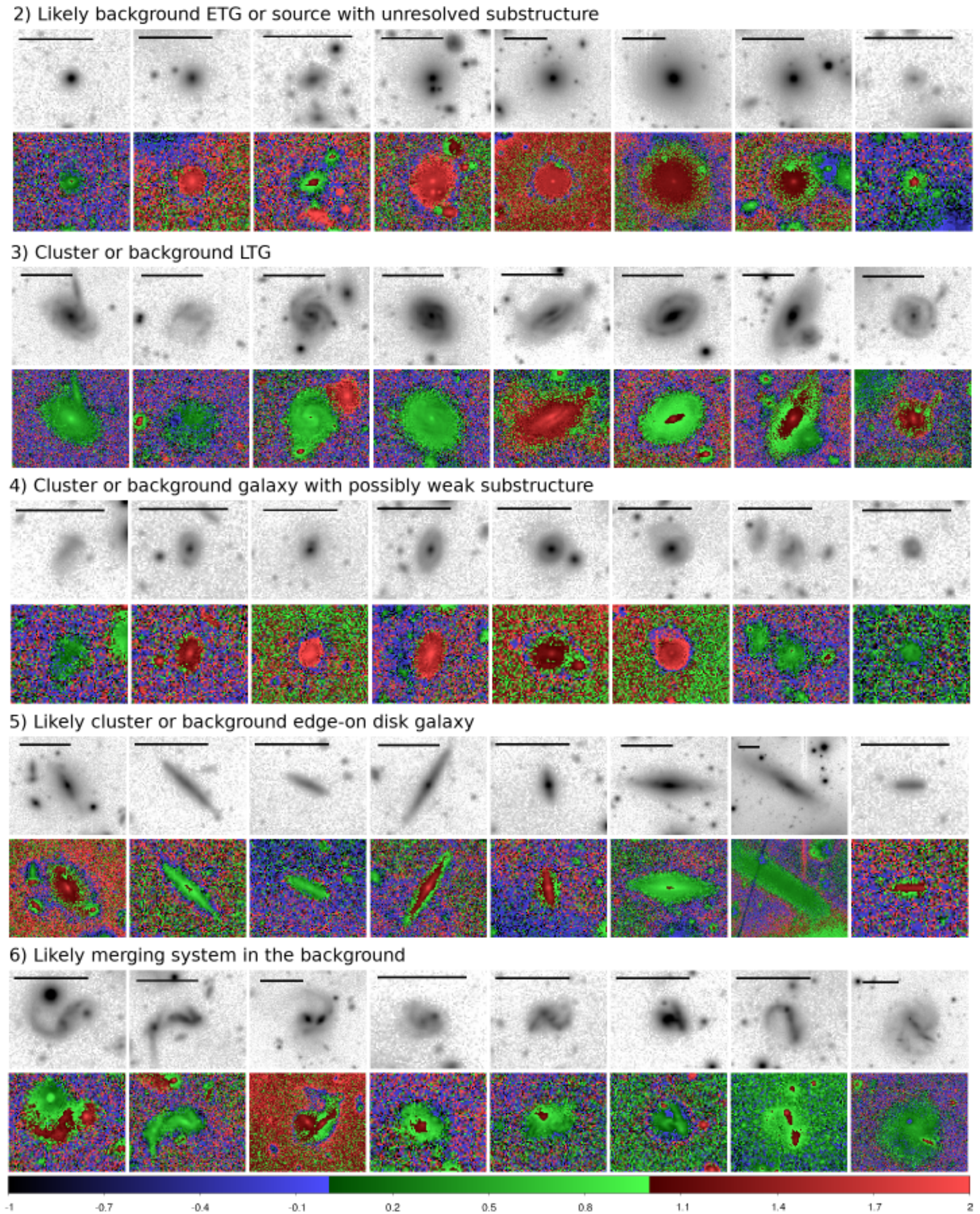}
    \caption{Sources in the direction of the Perseus cluster. Shown are the morphological categories  2) Likely background ETG or source with unresolved substructure; 3) Cluster or background LTG; 4) Cluster or background galaxy with possibly weak substructure; 5) Likely cluster or background edge-on disk galaxy; 6) Likely merging system in the background. For each source we display the {\it Subaru} $r$-band image in the top panel, and the $(g-r)_0$ color map in the bottom panel. The scale bar in each $r$-band image corresponds to a length of $10$\,arcsec. The color bar at the bottom of the figure indicates the $(g-r)_0$ values shown in the color maps.}
    \label{fig:fig6b}
\end{figure*}

\begin{figure*}
	\includegraphics[width=\textwidth]{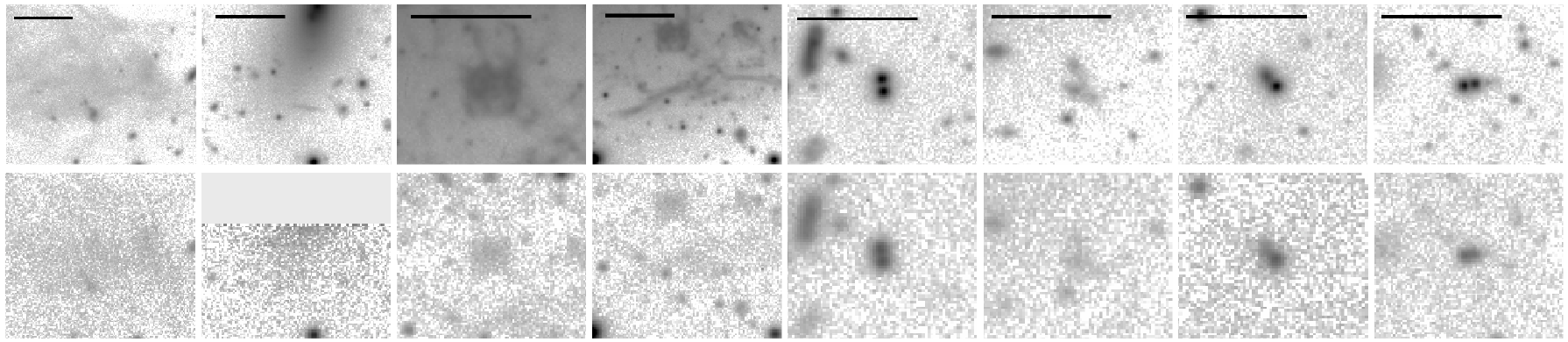}
    \caption{Excluded sources based on their appearance in the {\it Subaru} $r$-band images (top panels). For comparison, we show the {\it WHT} $V$-band images in the bottom panels. The scale bar in each $r$-band image corresponds to a length of $10$\,arcsec.}
    \label{fig:fig6c}
\end{figure*}

We provide an overview of the number of sources in each category in Table~\ref{tab:tab4}, and we show a selection of the classified sources in Figures~\ref{fig:fig6a}--\ref{fig:fig6c}. We discuss the described color substructure in a more quantitative way by measuring the color gradients in Figure~\ref{fig:figA2} in Appendix~\ref{sec:secA2}.

Figure~\ref{fig:fig7} illustrates the performance of our morphological classification in the color-magnitude diagram, highlighting the sample of dE/ETG cluster candidates. It can be seen that the dE/ETG candidates condense out as a tight red sequence, characteristic of an early-type cluster population \citep[e.g.,][]{Roediger2017}. The broad distribution of all other sources is a mixture of different categories, as illustrated by the image stamps in the figure. A detailed analysis of the Perseus cluster dE/ETG color-magnitude relation will be the subject of a forthcoming paper.

\begin{figure*}
	\includegraphics[width=\textwidth]{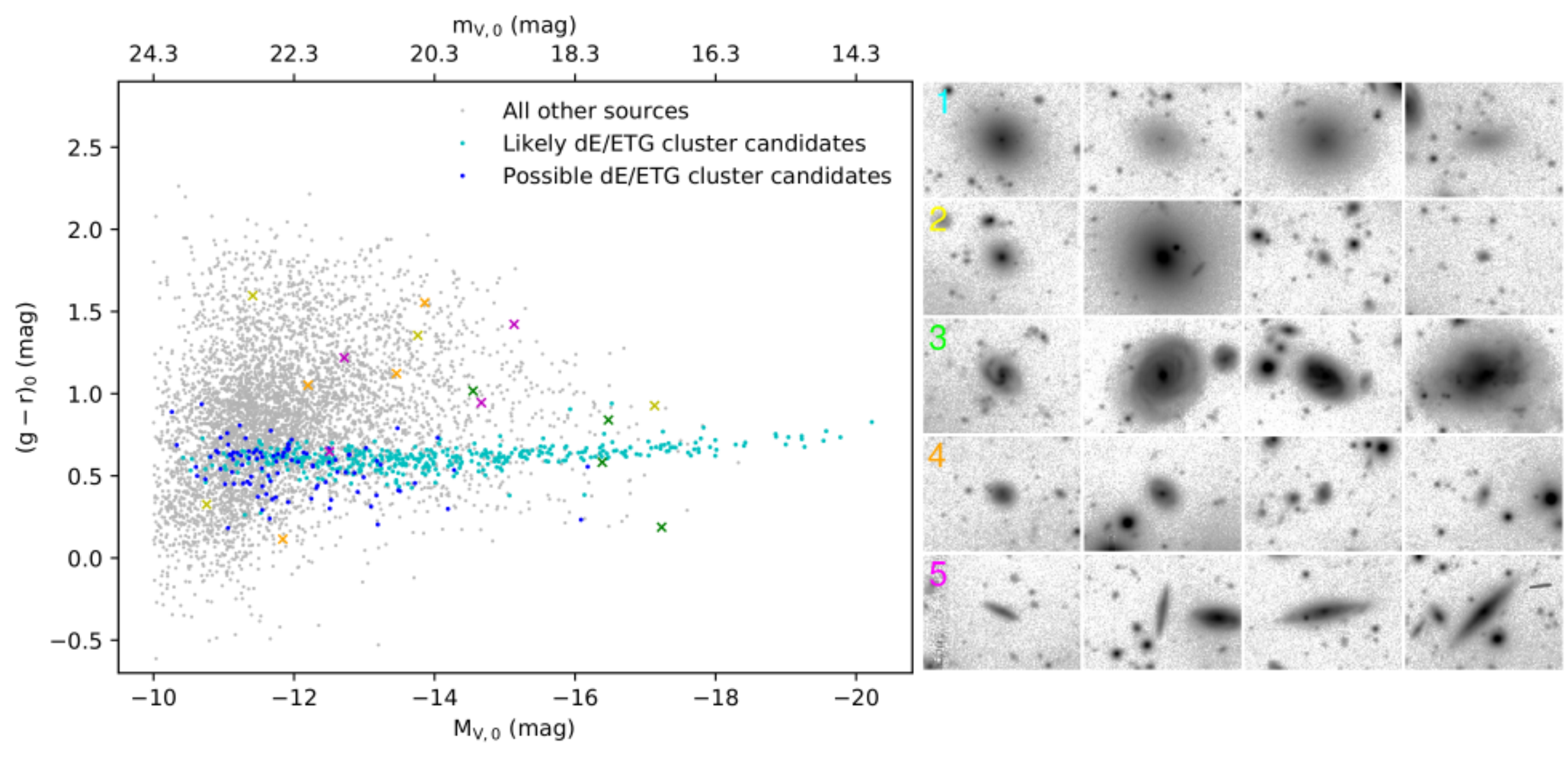}
    \caption{Left panel: color-magnitude diagram for the morphological categories (1) dE/ETG cluster candidates (highlighted with cyan and blue dots), (2) likely background ETGs or sources with unresolved substructure, (3) cluster or background LTGs, (4) cluster or background galaxies with possibly weak substructure, and (5) likely cluster or background edge-on disk galaxies. Right panel: images of the galaxies marked with the colored crosses in the left panel and some examples of the dE/ETG candidates. All images have a size of $25 \times 18$\,arcsec and show the {\it Subaru} $r$-band data. We show the ($r-z$)$_0$ versus $M_{V,0}$ diagram in Figure~\ref{fig:figA3} in Appendix~\ref{sec:secA3}.}
    \label{fig:fig7}
\end{figure*}

\subsection{Nucleation Classification of the dE/ETG Cluster Candidates} \label{sec:5.1}

We analyzed the dE/ETG cluster candidates for the presence of nuclei based on the {\it Subaru} $r-$band images. We classified a dE/ETG candidate as nucleated if we detected an unresolved high surface brightness point source centered on the respective galaxy. We identified nuclei in $182$ of the candidates. In the brighter luminosity regime, $34$ of the candidates show possible bulges or bright central sources that appear more extended compared to typical dE nuclei. For $38$ dE/ETG candidates, we could not unambiguously classify the source as nucleated or nonnucleated. Four sources show an accumulation of several brighter unresolved point sources near their center or slightly offset from it.\footnote{PCC 4867, PCC 2251, PCC 4304, PCC 5111.} The majority of the uncertain candidates show a central light concentration that could harbor a nucleus, but the potential nucleus does not stand out clearly as a distinct point source. In $242$ dE/ETG candidates, we could not identify a nucleus in our data and classified them as nonnucleated. We note, however, that nuclei could be missed in very faint sources or in more compact sources with a low brightness contrast between the nucleus and the galaxy main body.

\begin{figure*}
	\includegraphics[width=\textwidth]{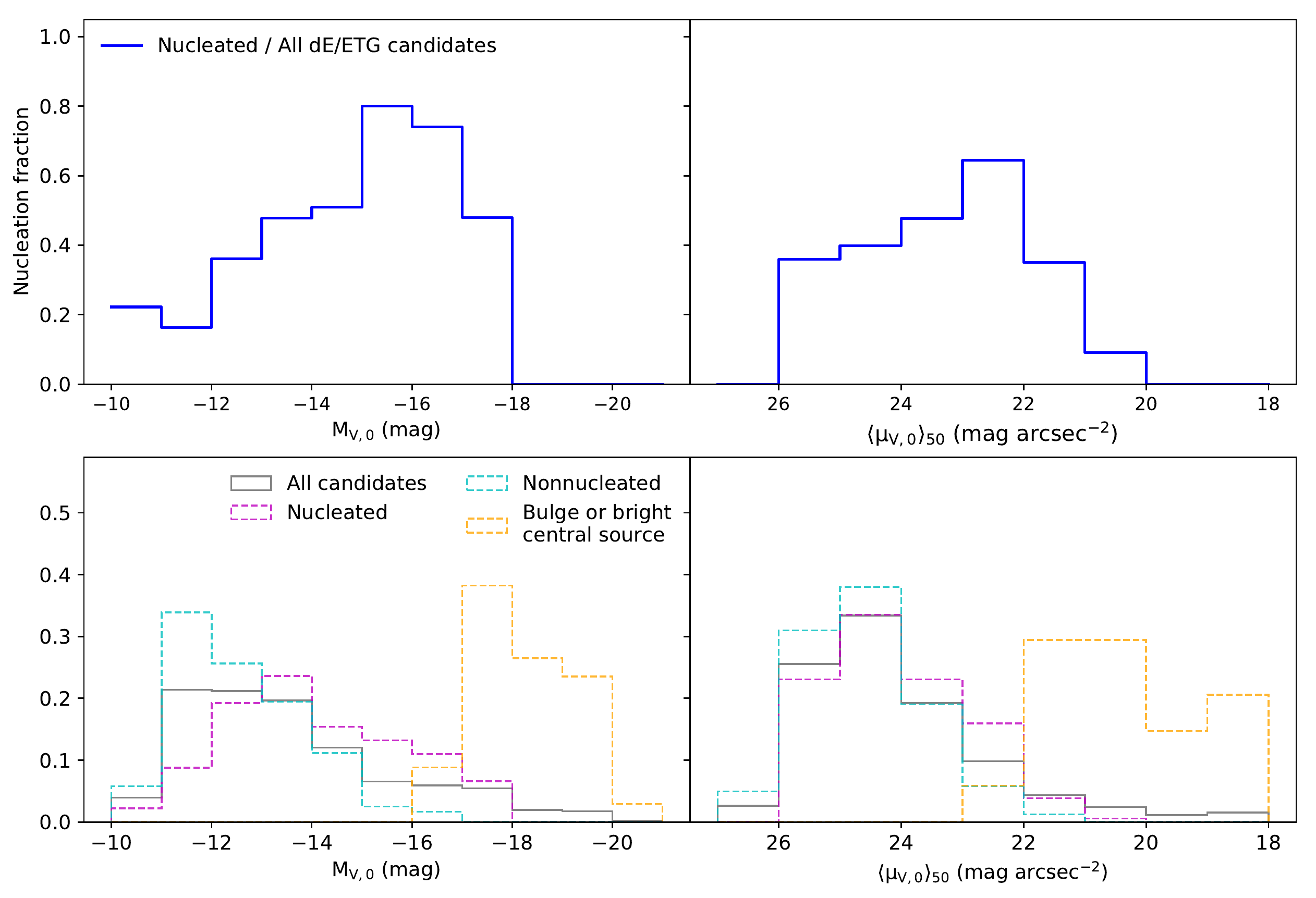}
    \caption{{\it Top panels:} nucleation fraction of Perseus dE/ETG candidates as a function of absolute magnitude or surface brightness, excluding the $38$ candidates with unsure nucleation classification. {\it Bottom panels:} luminosity distribution of nucleated and nonnucleated candidates and candidates with bulges or bright central sources in comparison to the entire dE/ETG subsample.}
    \label{fig:fig8}
\end{figure*}

Figure~\ref{fig:fig8} summarizes the nucleation properties of our sample. We found an average nucleation fraction of $47$\,per\,cent in the luminosity range $-10 \geq M_{V,0} \geq -18$\,mag when excluding the dE/ETG candidates with unsure nucleation classification. The nucleation fraction increases toward brighter luminosities and brighter surface brightnesses, with a peak at $M_{V,0} = -15$ to $-16$\,mag\footnote{We note that the nucleation fraction continuously increases as a function of luminosity when only considering the likely dE/ETG candidates.} and $\langle\mu_{V,0}\rangle_{50}= 22 - 23$\,mag\,arcsec$^{-2}$.  At brighter luminosities and surface brightnesses, the nucleation fraction drops again, due to an increase in the fraction of dE/ETG candidates harboring brighter central sources or bulges. The increase of the nucleation fraction with luminosity for faint ETGs is consistent with observations in other nearby clusters, like Fornax, Virgo, and Coma \citep{Sandage1985, Cote2006, denBrok2014, OrdenesBriceno2018b, Venhola2019}.

\section{The Catalog} \label{sec:sec6}

\begin{figure}
	\includegraphics[width=\columnwidth]{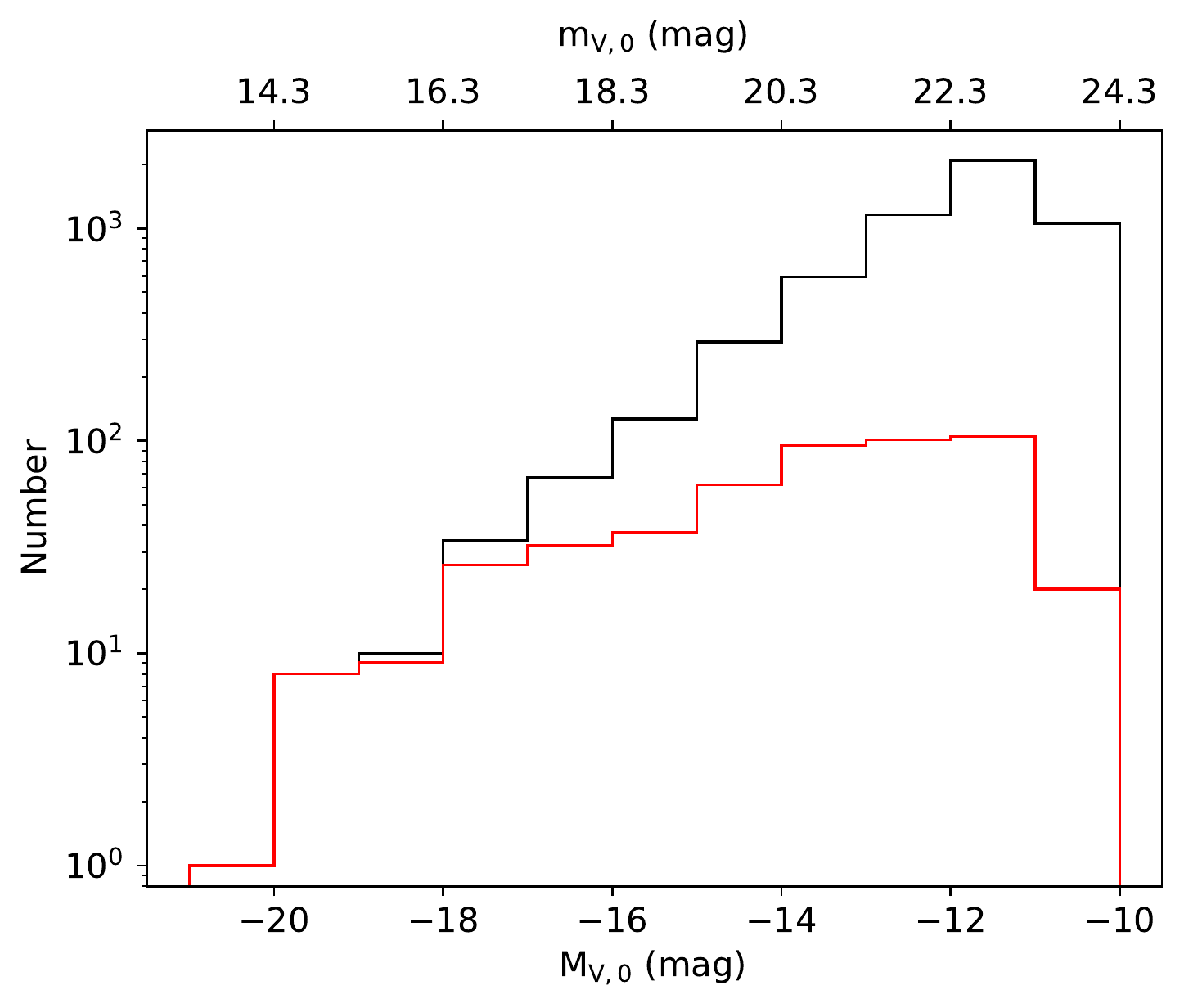}
    \caption{Luminosity distribution of the full PCC (black histogram) and the dE/ETG candidate subsample (red histogram).}
    \label{fig:LF}
\end{figure}

\begin{splitdeluxetable*}{lccccccccccBlcccccccccccc}
\tablecaption{Excerpt of the PCC Containing $5437$ Morphologically Classified Sources in the Direction of the Perseus Galaxy Cluster \label{tab:tab5}}  
\tabletypesize{\footnotesize}
\tablehead{
\colhead{ID} & \colhead{R.A. (J2000)} & \colhead{Decl. (J2000)} & \colhead{$m_{V,0}$} & \colhead{$r_{50}$} & \colhead{$\langle\mu_{V,0}\rangle_{50}$} & \colhead{n} & \colhead{$b/a$} & \colhead{$\theta$} & \colhead{$A_V$} & \colhead{pflag} &  \colhead{resflag} & \colhead{bgflag} & \colhead{(g-r)$_{0,aper1}$} & \colhead{(r-z)$_{0,aper1}$} & \colhead{(g-r)$_{0,aper2}$} & \colhead{(r-z)$_{0,aper2}$} & \colhead{(g-r)-flag}  & \colhead{(r-z)-flag}& \colhead{$A_g$} & \colhead{$A_r$} & \colhead{$A_z$} & \colhead{mflag} & \colhead{nuc}\\
& \colhead{(deg)} & \colhead{(deg)} & \colhead{(mag)} & \colhead{(arcsec)} & \colhead{(mag\,arcsec$^{-2}$)} &  &  & \colhead{(deg)} & \colhead{(mag)} &  &  &  & \colhead{(mag)} & \colhead{(mag)} & \colhead{(mag)} & \colhead{(mag)} & & & \colhead{(mag)} & \colhead{(mag)} & \colhead{(mag)} &  &
} 
\startdata
PCC 0001 & 49.2355 & +41.5722 & 20.51$\pm0.04$ & 1.44$\pm0.10$ & 20.80$\pm0.16$ & 1.07 & 0.10 & -64.5$\pm1.7$ & 0.41 & 5 & 0 & 0 & 0.69$\pm0.01$ & \phantom{-}0.25$\pm0.02$ & 0.63$\pm0.02$ & 0.28$\pm0.03$ & 0 & 0 & 0.50 & 0.35 & 0.19 & 5 & \nodata\\ 
PCC 0002 & 49.2366 & +41.4013 & 20.44$\pm0.04$ & 2.18$\pm0.14$ & 21.96$\pm0.15$ & 0.79 & 0.14$\pm0.02$ & \phantom{-}47.7$\pm1.3$ & 0.42 & 3 & 0 & 2 & 1.53$\pm0.02$ & \phantom{-}1.01$\pm0.02$ & 1.38$\pm0.04$ & 0.97$\pm0.03$ & 1 & 1 & 0.52 & 0.36 & 0.20 & 5 & \nodata\\
PCC 0003 & 49.2370 & +41.4336 & 21.58$\pm0.04$ & 1.11$\pm0.07$ & 22.41$\pm0.15$ & 0.07$\pm0.30$ & 0.28$\pm0.06$ & \phantom{-}58.4$\pm3.7$ & 0.41 & 1 & 0 & 0 & 1.25$\pm0.03$ & \phantom{-}0.71$\pm0.03$ & 1.19$\pm0.04$ & 0.67$\pm0.05$ & 0 & 0 & 0.50 & 0.35 & 0.19 & 5 & \nodata\\
PCC 0004 & 49.2377 & +41.5285 & 21.03$\pm0.06$ & 1.08$\pm0.10$ & 23.14$\pm0.21$ & 0.69$\pm0.16$ & 0.96$\pm0.07$ & \phantom{-}\nodata & 0.40 & 1 & 0 & 2 & 0.58$\pm0.03$ & \phantom{-}0.25$\pm0.07$ & 0.42$\pm0.06$ & 0.15$\pm0.14$ & 0 & 0 & 0.50 & 0.34 & 0.19 & 2 & \nodata\\
PCC 0005 & 49.2380 & +41.4346 & 20.27$\pm0.02$ & 1.24$\pm0.04$ & 21.94$\pm0.08$ & 0.80$\pm0.08$ & 0.48$\pm0.02$ & -47.5$\pm1.9$ & 0.41 & 1 & 0 & 0 & 0.70$\pm0.01$ & \phantom{-}0.50$\pm0.02$ & 0.53$\pm0.03$ & 0.48$\pm0.05$ & 0 & 0 & 0.50 & 0.35 & 0.19 & 5 & \nodata\\
PCC 0006 & 49.2388 & +41.4631 & 19.03$\pm0.00$ & 0.69$\pm0.01$ & 19.28$\pm0.02$ & 1.35$\pm0.06$ & 0.43$\pm0.01$ & \phantom{-}86.9$\pm0.6$ & 0.40 & 1 & 1 & 0 & 0.39$\pm0.00$ & -0.11$\pm0.01$ & \nodata & \nodata & 0 & 0 & 0.50 & 0.34 & 0.19 & 6 & \nodata\\ 
\enddata
\tablecomments{The columns list the following quantities: {\it ID}: identifier of the catalog source sorted by increasing R.A. {\it R.A.}: right ascension. {\it Decl.}: declination. {\it $m_{V,0}$}: apparent $V$-band magnitude corrected for extinction. {\it $r_{50}$}: half-light radius. {\it $\langle\mu_{V,0}\rangle_{50}$}: effective $V$-band surface brightness within the half-light radius corrected for extinction. {\it n}: S\'{e}rsic index. {\it $b/a$}: axis ratio. {\it $\theta$}: position angle measured north over east, where north is up and east is to the left. {\it $A_V$}: $V$-band extinction based on the reddening maps of \citet{Schlafly2011}. {\it pflag}: photometry processing flag given in Table~\ref{tab:tab3}. {\it resflag}: residual flag for sources fitted with \textsc{galfit}, where $resflag = 1$ flags sources with significant residuals after subtraction of the \textsc{Galfit} model, and  $resflag = 0$ denotes no obvious residuals. {\it bgflag}: background flag, where $bgflag = 0$ indicates that the local background was fitted and subtracted by \textsc{galfit}, $bgflag = 1$ denotes that a \textsc{SExtractor} generated background map with a \texttt{BACK\_SIZE} parameter of $256$\,pixels was subtracted, and $bgflag = 2$ refers to a subtracted \textsc{SExtractor} background map with a \texttt{BACK\_SIZE} parameter of $32$\,pixels. {\it (g-r)$_{0,aper1}$}: $g-r$ aperture color corrected for extinction measured within aperture 1, where $r_{aper1} = 1\,r_{50}$ for sources with $r_{50} > 4$\,pixels, and $r_{aper1} = 4$\, pixels for sources with $r_{50} \leq 4$\,pixels. {\it (r-z)$_{0,aper1}$}: $r-z$ aperture color measured within aperture 1 corrected for extinction. {\it (g-r)$_{0,aper2}$}: $g-r$ aperture color corrected for extinction measured only for sources with $r_{50} > 4$\,pixels within aperture 2, where $r_{aper2} = 2\,r_{50}$. {\it (r-z)$_{0,aper2}$}: $r-z$ aperture color measured within aperture 2 corrected for extinction. {\it (g-r)-flag}: flag indicating whether the source contains pixels of bleeding trails from bright stars within 3\,arcsec from the source center which may contaminate the color ({\it (g-r)-flag} = 1), or not ({\it (g-r)-flag} = 0).  {\it (r-z)-flag}: similar to {\it (g-r)-flag}, but with respect to (r-z). {\it $A_g$}: $g$-band extinction. {\it $A_r$}: $r$-band extinction. {\it $A_z$}: $z$-band extinction. $A_g$, $A_r$, and $A_z$ are based on the reddening maps of \citet{Schlafly2011}. {\it mflag}: morphology flag according to Table~\ref{tab:tab4}, where $mflag=11$ indicates dE/ETG candidates classified as {\it possible} candidates only. {\it nuc}: nucleation flag for sources with $mflag=1$ or $11$, where $nuc=1$ denotes nucleated sources, $nuc=2$ nonnucleated sources, $nuc=3$ unsure cases, and $nuc=4$ sources with a bulge or a bright central source. This table is available in its entirety in the electronic version of the paper.}
\end{splitdeluxetable*}

\begin{figure*}
\centering
	\includegraphics[width=0.6\textwidth]{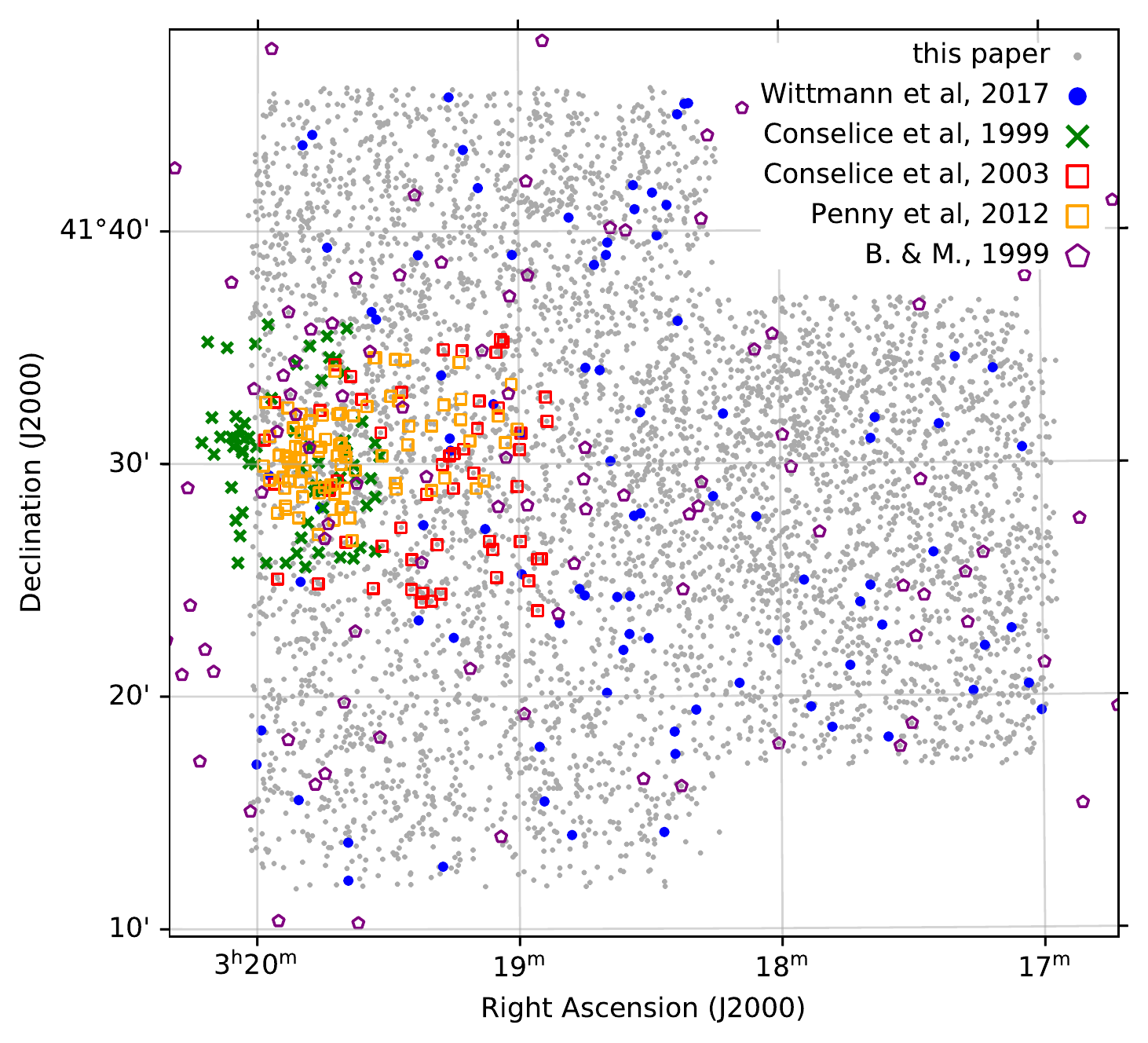}
    \caption{Coverage of the PCC (gray dots) in comparison to the Perseus cluster studies of \citet{Wittmann2017}, \citet{Conselice1999}, \citet{Conselice2003}, \citet{Penny2012}, and \citet{Brunzendorf1999}, with the corresponding symbols detailed in the legend.}
    \label{fig:catcomp}
\end{figure*}

\begin{splitdeluxetable*}{cccccccccBcccccccccBccccccc}
\tablecaption{Excerpt of the PCC Literature Crossmatch \label{tab:crossmatch}}
\tablehead{
\colhead{ID} & \colhead{R.A. (PCC)} & \colhead{Decl. (PCC)} & \colhead{LIT1} & \colhead{R.A. (LIT1)} & \colhead{Decl. (LIT1)} & \colhead{LIT2} & \colhead{R.A. (LIT2)} & \colhead{Decl. (LIT2)} & \colhead{LIT3} & \colhead{R.A. (LIT3)} & \colhead{Decl. (LIT3)} & \colhead{LIT4} & \colhead{R.A. (LIT4)} & \colhead{Decl. (LIT4)} & \colhead{LIT5} & \colhead{R.A. (LIT5)} & \colhead{Decl. (LIT5)} & \colhead{LIT6} & \colhead{R.A. (LIT6)} & \colhead{Decl. (LIT6)} & \colhead{NED} & \colhead{R.A. (NED)} & \colhead{Decl. (NED)}  & \colhead{$v$ (NED)}\\
& \colhead{(deg)} & \colhead{(deg)} & & \colhead{(deg)} & \colhead{(deg)} & & \colhead{(deg)} & \colhead{(deg)} & & \colhead{(deg)} & \colhead{(deg)} & & \colhead{(deg)} & \colhead{(deg)} & & \colhead{(deg)} & \colhead{(deg)} & & \colhead{(deg)} & \colhead{(deg)} & & \colhead{(deg)} & \colhead{(deg)} & \colhead{(km\,s$^{-1}$)}
} 
\startdata
PCC-0027  &  49.2475 & 41.3259 & \nodata & \nodata & \nodata & \nodata & \nodata & \nodata & \nodata & \nodata & \nodata & \nodata & \nodata & \nodata & \nodata & \nodata & \nodata & \nodata & \nodata & \nodata & 2MASS J03165939+4119332 & 49.248 & 49.248 & 39557\\
PCC-0040  &  49.2516 & 41.3224 & \nodata & \nodata & \nodata & \nodata & \nodata & \nodata & \nodata & \nodata & \nodata & \nodata & \nodata & \nodata & \nodata & \nodata & \nodata & 1 & 49.2516 & 49.2516 & \nodata & \nodata & \nodata & \nodata\\
PCC-0093  &  49.2636 & 41.3414 & \nodata & \nodata & \nodata & \nodata & \nodata & \nodata & \nodata & \nodata & \nodata & \nodata & \nodata & \nodata & \nodata & \nodata & \nodata & 2 & 49.2636 & 49.2636 & \nodata & \nodata & \nodata & \nodata\\ 
PCC-0124  &  49.2684 & 41.5109 & \nodata & \nodata & \nodata & \nodata & \nodata & \nodata & \nodata & \nodata & \nodata & \nodata & \nodata & \nodata & \nodata & \nodata & \nodata & 3 & 49.2684 & 49.2684 & \nodata & \nodata & \nodata & \nodata\\ 
\enddata
\tablecomments{Listed are the identifiers and coordinates of all PCC source that could be matched to at least one of the following literature sources. LIT1: identifier given in \citet{Conselice1999}. LIT2: identifier given in \citet{Brunzendorf1999}. LIT3: identifier given in \citet{Conselice2003}. LIT4: identifier given in \citet{Penny2008}. LIT5: identifier given in \citet{Penny2012}. LIT6: identifier given in \citet{Wittmann2017}. NED: identifier given in NED. The complete table is available in the electronic version of the paper. }
\end{splitdeluxetable*}

\begin{deluxetable*}{lllllll}
\tablecaption{Summary of the Latest Member Catalogs of the Perseus, Fornax, Virgo, and Coma Clusters \label{tab:catcomp}}
\tablecolumns{7}
\tablehead{
\colhead{Cluster} & \colhead{Focus} & \colhead{No. of} & \colhead{Magnitude Range}  & \colhead{Coverage} & \colhead{Completeness} & \colhead{Reference}\\
 &  & \colhead{Galaxies} & (mag) & (deg$^2$) &  & 
}
\startdata
Perseus & Faint ETGs & 496 & $-10.3 \geq M_V \geq -20.2$ & 0.3 & $M_V = -12$\,mag,  & (1)\\
 &  &  &  &  & $\langle\mu_{V,0}\rangle_{50} = 26$\,mag\,arcsec$^{-2}$\tablenotemark{a} & \\
Perseus & Faint LSBGs\tablenotemark{b} & 89 &  $-11.8 \geq M_V \geq -15.5$ & 0.3 & \nodata & (2)\\
Perseus & Faint ETGs & 53 & $-12.5 \geq M_B \geq -16.0$ & 0.05 & $M_B = -10.7$\,mag\tablenotemark{c} & (3), (4)\\
Fornax & Faint galaxies & 470 ETGs, & $-8.9 \geq M_r \geq -18.7$ & 26 & $M_r = -10.5$\,mag, & (5)\\
&&  94 LTGs &&&$\langle\mu_{V,0}\rangle_{50} = 26$\,mag\,arcsec$^{-2}$\tablenotemark{a} & \\
Fornax & Faint LSBGs\tablenotemark{d} & 205 & $-8.9 \geq M_r \geq -15.8$ & 4 & \nodata & (6)\\
Fornax & Faint galaxies & 384 & $-8.8 \geq M_i \geq -18.8$ & 9.5 & \nodata & (7)\\ 
Fornax & Faint galaxies & 258 & $-9.4 \geq M_i \geq -18.5$ & 3.1 & \nodata & (8)\\
Virgo & Galaxies & 380 ETGs, & $-7.4 \geq M_g \geq -22.2$ & 3.7 & $M_g = -9.1$\,mag, & (9), (10)\\
 &  & 24 LTGs &&&  $\langle\mu_g\rangle_{50} \simeq 27$\,mag\,arcsec$^{-2}$\tablenotemark{a} & \\
Coma & Faint LSBGs\tablenotemark{e} & 854 & $-10.9 \geq M_R \geq -16.9$ & 4.6 & \nodata & (11)\\
Coma & Faint ETGs & 200 & $-12.4 \geq M_{F814W} \geq -19.0$ & 0.08 & \nodata & (12)\\
Coma & Galaxies & 234 ETGs, & $-14.0 \geq M_B \geq -22.4$ & 0.6 & $M_B = -14$\,mag\tablenotemark{a} & (13)\\
 &  & 239 LTGs &&&&\\
\enddata
\tablenotetext{a}{50\,per\,cent completeness limit.}
\tablenotetext{b}{LSBGs = Low-surface brightness galaxies, with $\langle\mu_{V,0}\rangle_{50} \geq 24.8$\,mag\,arcsec$^{-2}$.}
\tablenotetext{c}{Luminosity distribution turnover.}
\tablenotetext{d}{with a central surface-brightness of $\mu_{r,c} \geq 23$\,mag\,arcsec$^{-2}$.}
\tablenotetext{e}{with $\mu_{R,c} \geq 22.4$\,mag\,arcsec$^{-2}$.}
\tablerefs{(1) Wittmann et al. 2019, (2) \citet{Wittmann2017}, (3) \citet{Conselice2002}, (4) \citet{Conselice2003}, (5) \citet{Venhola2018}, (6) \citet{Venhola2017}, (7) \citet{OrdenesBriceno2018a}, (8) \citet{Eigenthaler2018}, (9) \citet{SanchezJanssen2019}, (10) \citet{Ferrarese2016}, (11) \citet{Yagi2016}, (12) \citet{denBrok2014}, (13) \citet{Michard2008}.}
\end{deluxetable*}

Our final Perseus Cluster Catalog (PCC) includes the photometry and morphological classification of $5437$ sources in the direction of the Perseus cluster core. The catalog is available in the electronic version of the paper and the GAVO data center at \url{http://dc.g-vo.org/pcc/q/cone/form}, including $r$-band postage stamp images\footnote{{\it Subaru} $r$-band with a pixel scale of 0.168\,arcsec\,pixel$^{-1}$.} of the objects identified. We show an excerpt of the catalog in Table~\ref{tab:tab5}.

Figure~\ref{fig:LF} shows the luminosity distribution of the complete catalog, as well as the dE/ETG candidate subsample. Both distributions show a turnover in their number counts at $m_{V,0} = 22.3$\,mag ($M_{V,0} = - 11$\,mag), illustrating that the catalog is not complete at luminosities fainter than this. We note that the completeness estimate with the inserted model galaxies (see Section~\ref{sec:sec3.3}) indicates a 50\,per\,cent completeness limit already at $M_{V,0} = - 12$\,mag.

Our catalog is the largest and deepest catalog with a coherent coverage of $0.3$\,deg$^2$ that exists to date in the direction of the Perseus galaxy cluster core. Compared to the imaging survey of faint galaxies in the Perseus core region presented in \citet{Conselice2002, Conselice2003}, our data cover an area larger by roughly a factor of six, and have a turnover magnitude in the luminosity distribution fainter than about $0.5$\,mag. This allowed us to identify almost $10$ times more low-mass galaxy candidates. Figure~\ref{fig:catcomp} shows the coverage of our catalog in comparison to the Perseus cluster imaging studies of \citet{Conselice1999, Brunzendorf1999, Conselice2003, Penny2012} and \citet{Wittmann2017}. We provide the IDs of all PCC sources that were previously cataloged in Table~\ref{tab:crossmatch}, as well as available radial velocity measurements that we compiled from NED (also see Section~\ref{sec:sec8.2}).

The PCC is clearly deeper than the well-established Virgo Cluster Catalog \citep[VCC;][]{Binggeli1985} and the Fornax Cluster Catalog \citep[FCC;][]{Ferguson1989} with regard to their limiting magnitude of $m_{B_T} = 18$\,mag. However, both catalogs cover a much larger area of the respective cluster, reaching even beyond the virial radius, whereas our catalog only covers about one-third of the Perseus cluster virial radius. 

Compared to more recent imaging surveys, like the FDS Fornax cluster dwarf galaxy catalog \citep{Venhola2018}, which covers the Fornax cluster to its virial radius, our catalog reaches a similar 50\,per\,cent completeness limit with regard to surface brightness. The 50\,per\,cent completeness limit is, however, shallower by more than one magnitude with regard to luminosity. This is likely explained by the faint magnitude cut we apply to define our working sample (see Section~\ref{sec:sec3.2}). We provide a summary of the latest member catalogs focusing on faint galaxies of the Perseus, Fornax, Virgo, and Coma clusters in Table\,\ref{tab:catcomp}.

\section{Parameter Distributions of the Morphological Subclasses} \label{sec:sec7}

A morphological analysis, together with color information, provides a valuable tool to disentangle cluster dE/ETG candidates from late-type and background galaxies. If, however, one only has a data set at hand without subarcsecond resolution and lacking color information, the question is how large the contamination through noncluster sources will be in a certain parameter range. In the following, we use our morphologically categorized sample to investigate the $m_{V,0}-\langle\mu_{V,0}\rangle_{50}-r_{50}-n$ parameter distributions of the different subsamples, and discuss which parameter ranges yield the highest fraction of cluster dE/ETG candidates.

\vspace{1cm}

\subsection{The $m_{V,0}-r_{50}-\langle\mu_{V,0}\rangle_{50}$ Distribution} \label{sec:sec7.1}

\begin{figure*}
	\includegraphics[width=\textwidth]{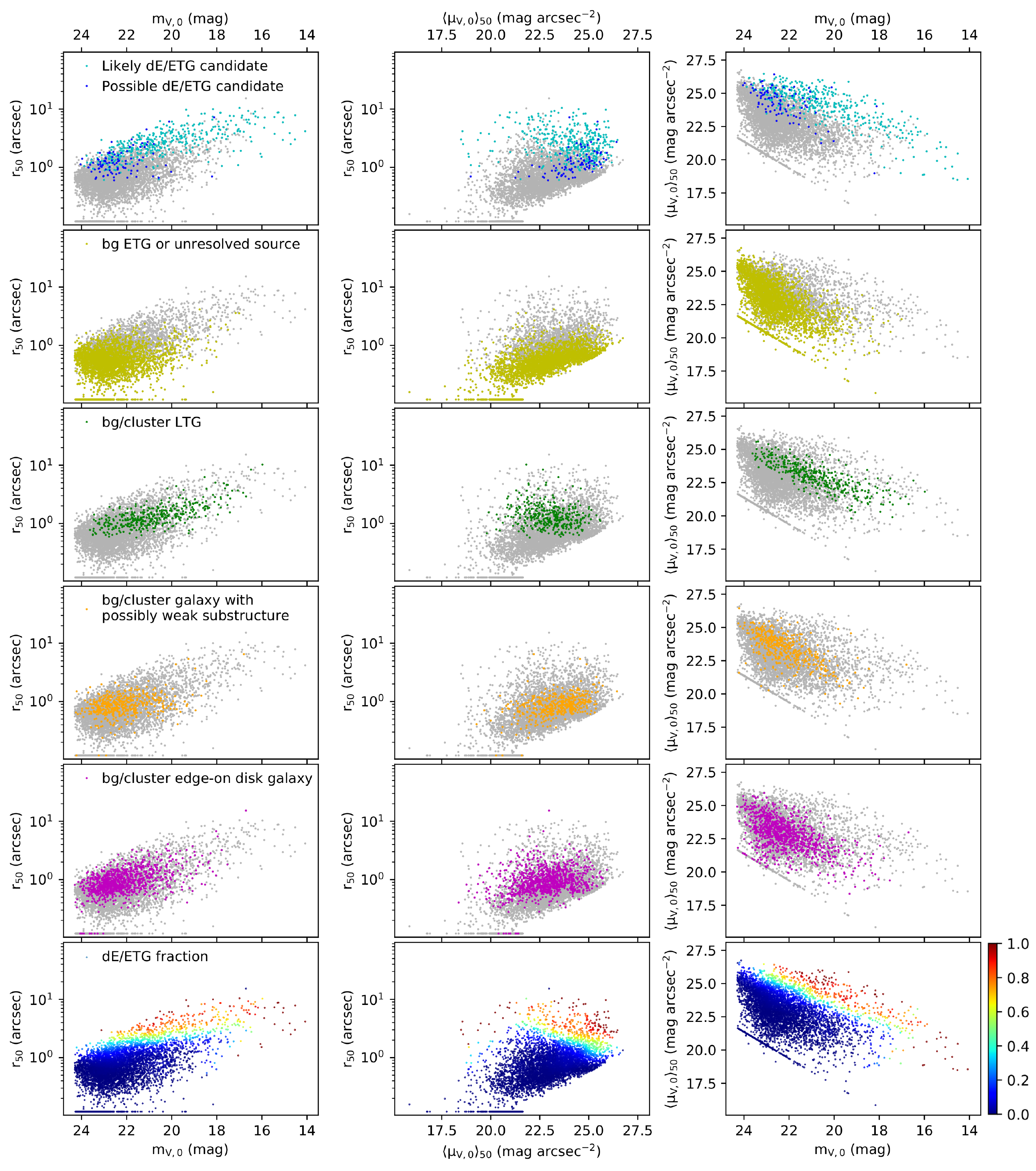}
    \caption{The $m_{V,0}-r_{50}-\langle\mu_{V,0}\rangle_{50}$ distribution of the morphological categories 1-5, defined in Table~\ref{tab:tab4}. In each row, we highlight one subclass and show the respective other sources as gray symbols. The last row displays the fraction of dE/ETG candidates in the respective parameter range. We calculated the dE/ETG fraction at the $M_{V,0}, r_{50}, \langle\mu_{V,0}\rangle_{50}$ position of each source in intervals of $m_{V,0} \pm 0.5$\,mag, $\langle\mu_{V,0}\rangle_{50} \pm 0.5$\,mag\,arcsec$^{-2}$ and $log(r_{50}) \pm 0.3$. The label ``bg'' in the legends refers to ``background''.}
    \label{fig:fig9}
\end{figure*}

Figure~\ref{fig:fig9} shows the $m_{V,0}-r_{50}-\langle\mu_{V,0}\rangle_{50}$ distribution of the morphological subclasses $1-5$, defined in Table~\ref{tab:tab4} as well as the dE/ETG candidate fraction with respect to all other sources. It can be seen that the highest dE/ETG fraction, and thus lowest contamination through other sources, is reached at the largest sizes at a given luminosity or surface brightness and at the brightest luminosities. For sources with $m_{V,0} > 16$\,mag, the dE/ETG fraction is, on average, above $50$\,per\,cent for sources with $r_{50} \gtrsim 1.6$\,arcsec, with an increasing dE/ETG fraction toward larger sizes. For sources with $m_{V,0} < 16$\,mag, the dE/ETG fraction is close to $100$\,per\,cent. We note that the fraction of the dE/ETG candidates classified as possible candidates increases toward fainter luminosities and smaller sizes, likely reflecting a more uncertain morphological classification in this parameter regime.

At the smallest sizes, below $r_{50} \sim 1$\,arcsec, sources from category $2$ dominate, including unresolved sources where the morphological characterization breaks down. The larger sources are likely background ellipticals. A few sources in the parameter range $M_{V,0} \lesssim -14$\,mag ($m_{V,0} = 20.3$\,mag) and $r_{50} \gtrsim 100$\,pc ($\widehat{=}\,0.3$\,arcsec) may, however, also be cluster compact elliptical galaxies \citep[e.g.,][]{SmithCastelli2008, Norris2014}. Among the faint and small sources, a few could be Perseus cluster ultracompact dwarf galaxies \citep{Penny2012, Penny2014a}. Indeed, six of our sources were identified as ultracompact dwarf candidates by \citet{Penny2012} based on {\it HST} imaging data. The sources of categories $3-5$, including LTGs, galaxies with possibly weak substructure, and edge-on disk galaxies, are mainly found at the intermediate size range. Some of these might be true Perseus cluster members. However, for these categories, it is not possible to discriminate between cluster and background galaxies based on morphology and color alone, requiring spectroscopic confirmation.

\subsection{The $M_{V,0}-$S\'{e}rsic Distribution} \label{sec:sec7.2}

It is known that there exists a well-defined relationship between galaxy luminosity and S\'{e}rsic index, where brighter galaxies have a higher S\'{e}rsic index \citep[e.g.,][]{Jerjen1997, Graham2003, Gavazzi2005}. For faint galaxies, various studies showed that this relation breaks down, leveling off to a constant value of S\'{e}rsic index $n \sim 1$ \citep[e.g.,][]{deRijcke2009, Misgeld2009, Lieder2012}. 

\begin{figure*}
	\includegraphics[width=\textwidth]{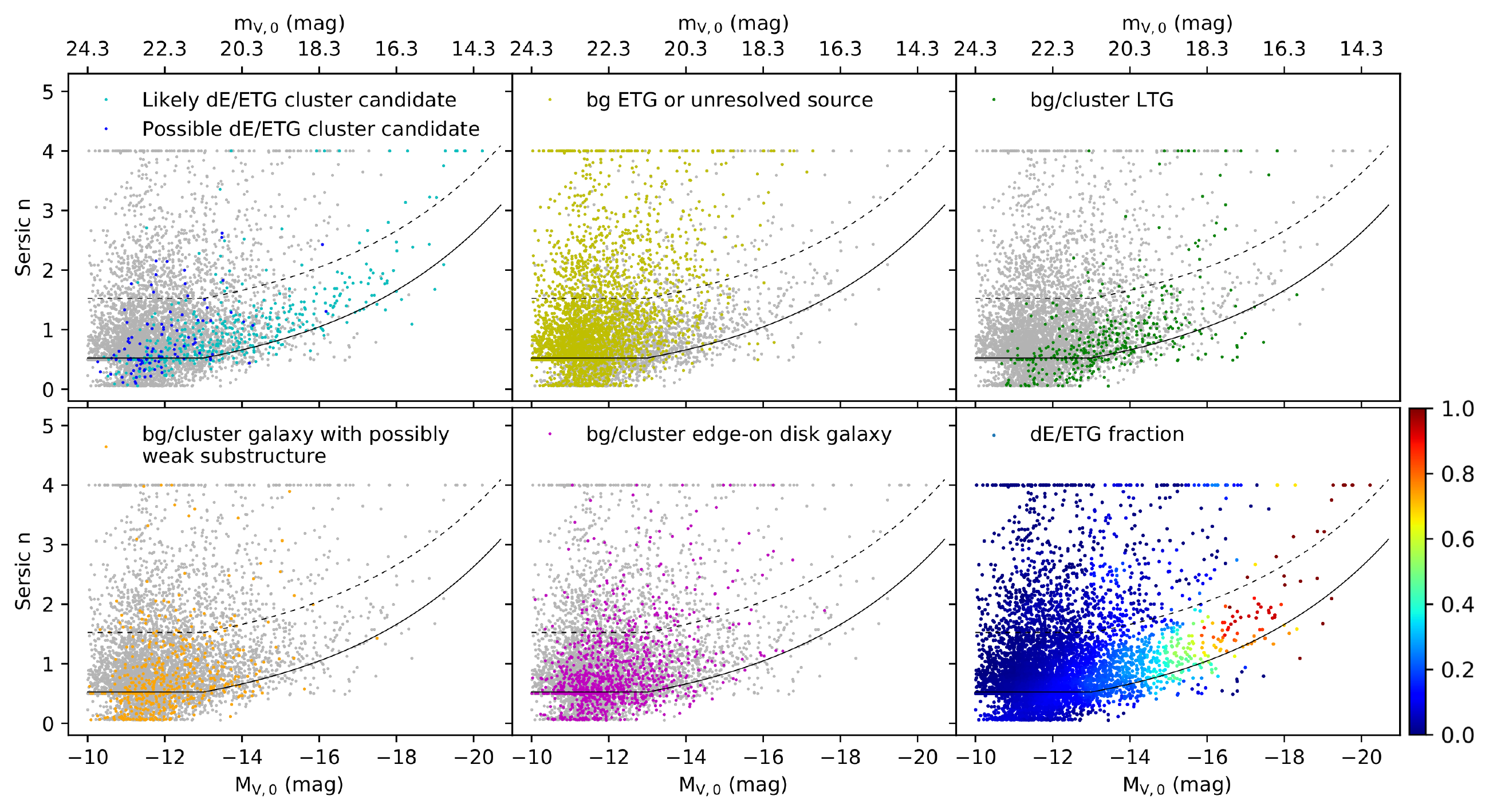}
    \caption{The $M_{V,0}-$S\'{e}rsic distribution of the morphological subclasses 1-5, defined in Table~\ref{tab:tab4}. In each panel, we highlight one subclass and show the respective other sources as gray symbols. The solid black curve in each panel shows the $M_{V,0} - n$ relation from \citet{Graham2003} for $M_{V,0} \leq -13$\,mag. For fainter magnitudes, the value of $n$ at $M_{V,0} = -13$\,mag is adopted (see Section~\ref{sec:sec7.2}). The dashed black curve indicates the $n + 1$ interval with respect to the $M_{V,0} - n$ relation. The bottom right panel displays the dE/ETG fraction at the $M_{V,0} ,n$ position of each source in intervals of $M_{V,0} \pm 0.5$\,mag and $n \pm 0.25$. The label ``bg'' in the legends refers to ``background''.}
    \label{fig:fig10}
\end{figure*}

In Figure~\ref{fig:fig10} we display the $M_{V,0}-n$ distribution of the morphological subclasses $1-5$, together with the literature $M_{V,0}-n$ relation. One can see that the vast majority of the sources classified as dE/ETG cluster candidates follow the $M_{V,0}-n$ relation. At bright luminosities ($M_{V,0} < -14.6$\,mag), the dE/ETG fraction is above $50$\,per\,cent within an interval of $n + 1$ from the literature relation, with an increasing dE/ETG fraction toward brighter luminosities. At fainter luminosities, however, the contamination by other sources increases dramatically.

A relation between S\'{e}rsic index and luminosity is also seen for the sources classified as LTGs. This might also indicate that a significant fraction of these are cluster members rather than background galaxies. The sources of the other three morphological categories show no $M_{V,0}-n$ relation, likely indicating that they are distributed across a wide range of redshifts.

\section{Literature Comparisons} \label{sec:sec8}

\subsection{{\it HST} Data Comparison for the dE/ETG Cluster Candidates} \label{sec:sec8.1}

We visually examined the sources classified as dE/ETG cluster candidates in archival {\it HST} imaging data in order to check whether we missed intrinsic substructure. In total, we checked eleven {\it HST} ACS/WFC3 frames, including $70$ of our candidates. The vast majority (59) show a smooth dE-like appearance in the {\it HST} images. Five candidates are only barely seen, and four are undetected. Two candidates look suspicious and might be in the background, although clear substructure is not revealed. The generally positive comparison with the {\it HST} data further supports our classification of the sources as dE/ETG candidates. We summarize the {\it HST} comparison in Table~\ref{tab:tabA4} in Appendix~\ref{sec:secA4}.

\subsection{Spectroscopically Confirmed Sources and Background Source Estimation} \label{sec:sec8.2}

We checked which sources in our catalog have available radial velocity measurements. For this, we used NED to compile all sources classified as galaxies that fall on our {\it WHT} mosaic footprint (see Section~\ref{sec:sec6}). We identified $49$ sources that match a source in our catalog. Of the matched sources, $28$ have velocities within $2\,\sigma$ of the Perseus cluster mean velocity $v_{Perseus} = 5370$\,km\,s$^{-1}$ \citep{Struble1999}, where $\sigma = 1300$\,km\,s$^{-1}$ \citep{Kent1983}. All of these are in the category of likely dE/ETG cluster candidates, thus confirming our morphological classification. Two sources have radial velocities within $2 - 3\,\sigma$ of the cluster mean velocity, of which one source is classified as a likely dE/ETG candidate and the other as a likely background ETG or unresolved source. 

The remaining $19$ sources are not cluster members, having $ v > v_{Perseus} + 3 \sigma$\,km\,s$^{-1}$. They fall into the morphological categories of likely background ETGs or unresolved sources (seven sources), cluster or background galaxies with late-type morphology (five sources), cluster or background galaxies with possible weak substructure (four sources), and cluster and background edge-on disk galaxies (one source). Two sources\footnote{PCC $314$ and PCC $1628$.} were morphologically classified as dE/ETG candidates. They are likely more distant ETGs, although they are characterized by a smooth morphology and lie close to the red sequence in the color-magnitude diagram, with ($g-r$)$_0 = 0.7$ and $0.8$\,mag, respectively.

From the comparison of our visual classification for sources with radial velocity measurements, we can also get a rough estimate of the expected contamination through background galaxies in our dE/ETG candidate sample. Of the sources visually classified as dE/ETG candidates, 28 are confirmed cluster members, whereas two are confirmed background systems. This results in a contamination rate of seven per\,cent. Thus, 35 of the 496 sources classified as dE/ETG candidates might be background contaminants.

We also performed an order-of-magnitude comparison to the background galaxy number density for Perseus estimated by \citet[][see their Appendix A4]{Weinmann2011}. The authors derived a background galaxy number density of 45\,Mpc$^{-2}$ for galaxies with $-16.7 \geq M_r \geq -19$\,mag by extrapolating the radial number density profile based on SDSS data. Considering sources in the same magnitude regime from our catalog that are not classified as dE/ETG candidates and thus should be dominated by background sources, we derive an average number density of 70\,Mpc$^{-2}$ within 700\,kpc. The value, higher by a factor of 1.6 compared to the one in \citet{Weinmann2011}, might be caused by cluster LTGs contaminating our background galaxy number estimate, since we did not visually distinguish between cluster and background systems for LTGs. The order-of-magnitude agreement, however, further strengthens our performed visual classification approach.

\subsection{Faint Low Surface Brightness Galaxies} \label{sec:sec8.3}

\subsubsection{Visual versus Automated Detection} \label{sec:sec8.3.1}

\citet{Wittmann2017} established a sample of faint low surface brightness galaxy candidates in the Perseus cluster core based on a visual detection in the {\it WHT} $V$-band data, which we also used in this work. In the following, we discuss how many of the visually identified sources were also detected with the automated detection method applied in this study. We found that $77$ of the $89$ faint low surface brightness galaxy candidates were successfully detected with our automated approach.\footnote{We note that two of them were excluded in this study, since one was revealed as cirrus and the other is only barely visible in the {\it Subaru} $r$-band data. These are the sources ID=31 and ID=57, which are listed as possible low surface brightness galaxy candidates in \citet{Wittmann2017}.} The $12$ missed sources  are all faint, with $M_{V,0} > -14$\,mag, and occupy the low surface brightness regime, with nine sources having $\langle\mu_{V,0}\rangle_{50} > 26$\,mag\,arcsec$^{-2}$. 

This demonstrates that our automated \textsc{SExtractor}-based source detection generally works well to extract  sources with $\langle\mu_{V,0}\rangle_{50} \lesssim 26$\,mag\,arcsec$^{-2}$ from the {\it WHT} data. For fainter surface brightness sources, however, a visual detection seems more reliable. This is also expected, given that our completeness estimate indicates that we miss more than $50$\,per\,cent of all sources with  $\langle\mu_{V,0}\rangle_{50} > 26$\,mag\,arcsec$^{-2}$ (see Section~\ref{sec:sec3.3}). In contrast, according to the completeness estimate given in \citet[][see their Figure\,2]{Wittmann2017}, the detection rate of visually identified sources with $\langle\mu_{V,0}\rangle_{50} = 26 - 27 $\,mag\,arcsec$^{-2}$ is still in the range of $70-90$\,per\,cent.

\subsubsection{New Ultradiffuse Galaxy Candidates} \label{sec:sec8.3.2}

\begin{figure}
	\includegraphics[width=\columnwidth]{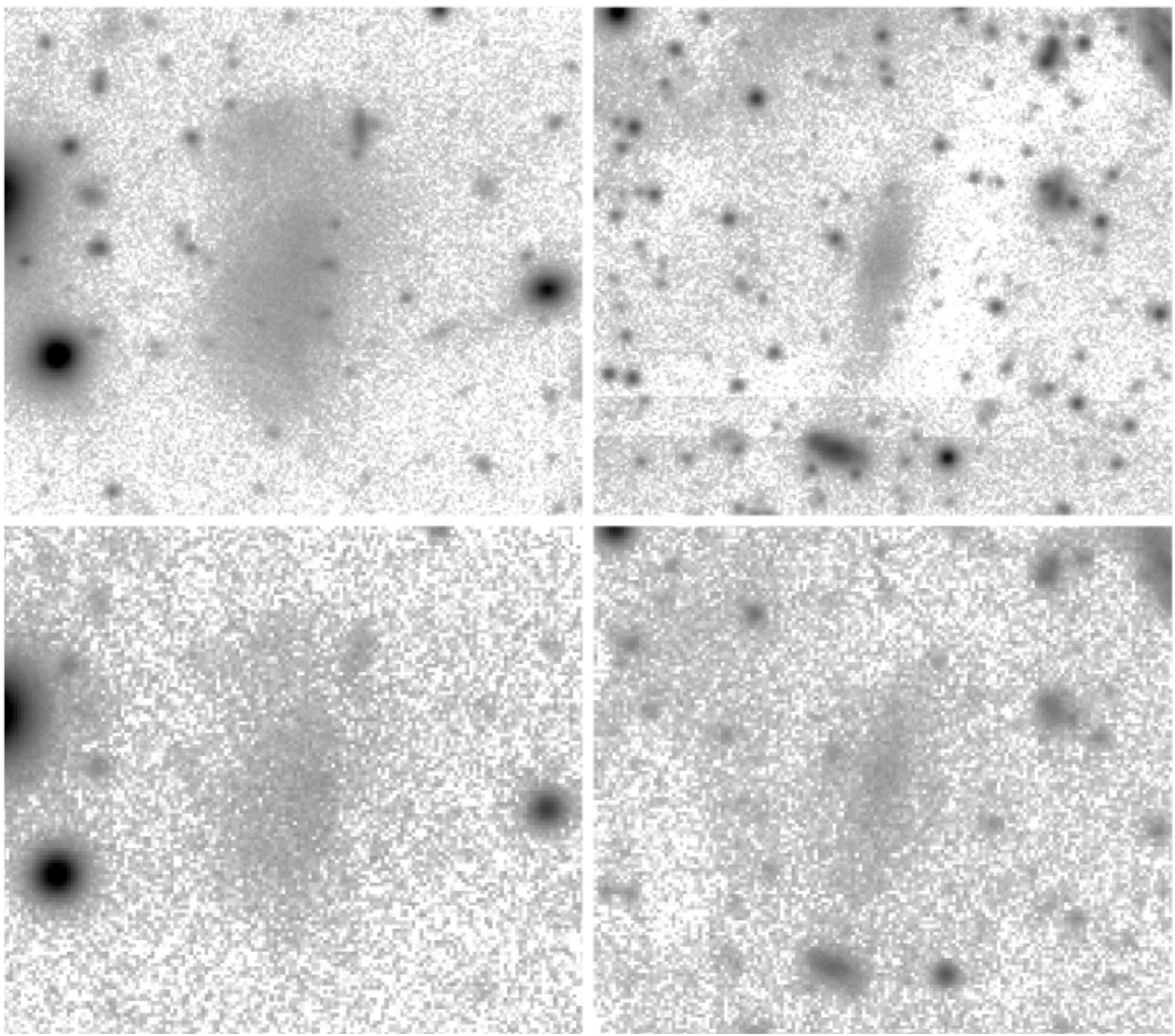}
    \caption{New ultradiffuse galaxy candidates. The top panels show the candidates PCC 2262 (left) and PCC 4017 (right) in the {\it Subaru} $r$-band data. In the top right panel, we additionally subtracted a \textsc{SExtractor}-generated background map with a BACK\_SIZE of $45$\,pixels in order to reveal the candidate in the extended halo of a bright neighbor galaxy. The bottom panels show both candidates in the {\it WHT} $V$-band data. The width of each panel corresponds to $38$\,arcsec.}
    \label{fig:UDGs}
\end{figure}

Our dE/ETG sample also includes two candidates in the parameter regime $\langle\mu_{V,0}\rangle_{50} \geq 25$\,mag\,arcsec$^{-2}$ and $r_{50} \geq 1.5$\,kpc that fall into the category of ultradiffuse galaxy candidates and that were not previously reported. We note, however, that both objects are classified as possible candidates only, due to their not entirely smooth morphological appearance. We show both candidates in Figure~\ref{fig:UDGs}.

\section{Summary} \label{sec:sec9}

In this study, we established a new catalog of sources in the direction of the Perseus galaxy cluster core, with a focus on extracting low-mass galaxy cluster members. Our catalog covers an area of $0.3$\,deg$^2$ and includes $5437$ sources, among them $496$ early-type low-mass galaxy candidates that were selected based on their morphological appearance. The catalog reaches its $50$\,per\,cent completeness limit at $M_{V,0} = -12$\,mag and $\langle\mu_{V,0}\rangle_{50} = 26$\,mag\,arcsec$^{-2}$. This makes it to the largest and deepest catalog with coherent coverage compared to previous imaging studies of the Perseus galaxy cluster.

The source detection and photometry is based on {\it WHT} $V$-band imaging data covering $0.7 \times 0.7$\,Mpc$^2$ ($0.58 \times 0.58$\,deg$^2$) of the cluster core region. We used \textsc{SExtractor} to carry out an automated source detection and derived photometry for all detected sources with \textsc{Galfit}.

To perform the morphological classification and to measure aperture colors, we reduced archival {\it Subaru} $g-,r-,$ and $z$-band data. We used the $r$-band data for the morphological analysis, where we classified the sources into (1) dE/ETG cluster candidates, (2) likely background ETGs or sources with unresolved substructure, (3) cluster or background LTGs, (4) cluster or background galaxies with possibly weak substructure, (5) cluster or background edge-on disk galaxies, and (6) likely background merging systems.

We found that the dE/ETG cluster candidates form a tight red sequence in the color-magnitude diagram and follow the literature relation between absolute luminosity and S\'{e}rsic index, strengthening our morphological classification as early-type cluster members. The sources of the remaining morphological classes show a broad scatter in their parameter distributions, indicative of many of them being in the background of Perseus, at a wide range of redshifts. 

We classified the dE/ETG candidates as nucleated or nonnucleated galaxies and confirmed the trend of increasing nucleation fraction toward brighter luminosity or higher surface brightness in the regime of $M_{V,0} \geq -16$\,mag and $\langle\mu_{V,0}\rangle_{50} \geq 23$\,mag\,arcsec$^{-2}$, which is also seen for the early-type populations in other nearby galaxy clusters. At even higher luminosities and surface brightnesses, the nucleation fraction quickly declines due to the increasing fraction of galaxies harboring bulges.

In the $m_{V,0}-r_{50}-\langle\mu_{V,0}\rangle_{50}-n$ parameter space, the dE/ETG candidates show a significant overlap with the other morphological classes. We found, however, that the dE/ETG fraction is above $50$\,per\,cent in the largest size range of $r_{50} \gtrsim 1.6$\,arcsec, and along the literature $M_{V,0} - n$ relation for $M_{V,0} < -14.6$\,mag. This information may be useful for studies using data sets where a morphological classification is limited due to poor resolution of the data or for wide-field surveys covering large fractions on the sky, where a morphological analysis might be too time-consuming.

\acknowledgments
This work is based on data acquired at the {\it William Herschel Telescope}, which is operated on the island of La Palma by the Isaac Newton Group of Telescopes in the Spanish Observatorio del Roque de los Muchachos of the Instituto de Astrof\'{i}sica de Canarias (program 2012B/045). We thank Daniel Bialas for his help with the observations, as well as Simone Weinmann and Stefan Lieder for useful comments when preparing the {\it WHT} observing proposal.
This work is based in part on data collected at the Subaru Telescope and obtained from the SMOKA, which is operated by the Astronomy Data Center, National Astronomical Observatory of Japan.
We gratefully acknowledge the investigators proposing and executing the Subaru observations, Nozomu Tominaga (Subaru project S14B-048) and Toshifumi Futamase (S14B-030) and their co-observers Okabe-san, Hada-san, and Morokuma-san, as well as the staff of the Subaru/SMOKA help-desk for their fast and helpful support at several points throughout the work with the Subaru data.
We thank Markus Demleitner for his help with the GAVO data release.
C.W.\ is supported by the Deutsche Forschungsgemeinschaft (DFG; German Research Foundation) through project 394551440.
R.K.\ gratefully acknowledges partial funding support from the National Science Foundation under project AST-1664362 and from the National Aeronautics and Space Administration under project 80NSSC18K1498.
This research has made use of the NASA/IPAC Extragalactic Database (NED), which is operated by the Jet Propulsion Laboratory, California Institute of Technology, under contract with the National Aeronautics and Space Administration.
\software{HSC pipeline (Version 5.4), \textsc{SWarp} \citep{Bertin2002}, \textsc{SExtractor} \citep{Bertin1996}, \textsc{iraf} \citep{Tody1986, Tody1993}, \textsc{galfit} \citep{Peng2002, Peng2010}}.

%






\appendix

\section{Petrosian Photometry Uncertainties}
\label{sec:secA1}

We give an estimate of typical Petrosian photometry uncertainties based on model galaxies inserted into the {\it WHT} $V$-band data. For the estimate, we used the inserted model galaxies that we generated for the \textsc{SExtractor} parameter tuning and the completeness estimate (see Section~\ref{sec:sec3.1} and \ref{sec:sec3.3}), selecting models in the parameter range $M_{V,0} = -10$ to $-19$\,mag, $r_{50} = 0.2-5.0$\,kpc, and $\langle\mu_{V,0}\rangle_{50} = 19-27$\,mag\,arcsec$^{-2}$. We derived Petrosian photometry in the same way as for our working sample of real detected sources described in Section~\ref{sec:sec4.1} and compared the intrinsic to the measured model parameters.

\begin{figure}[h!]
	\includegraphics[width=\textwidth]{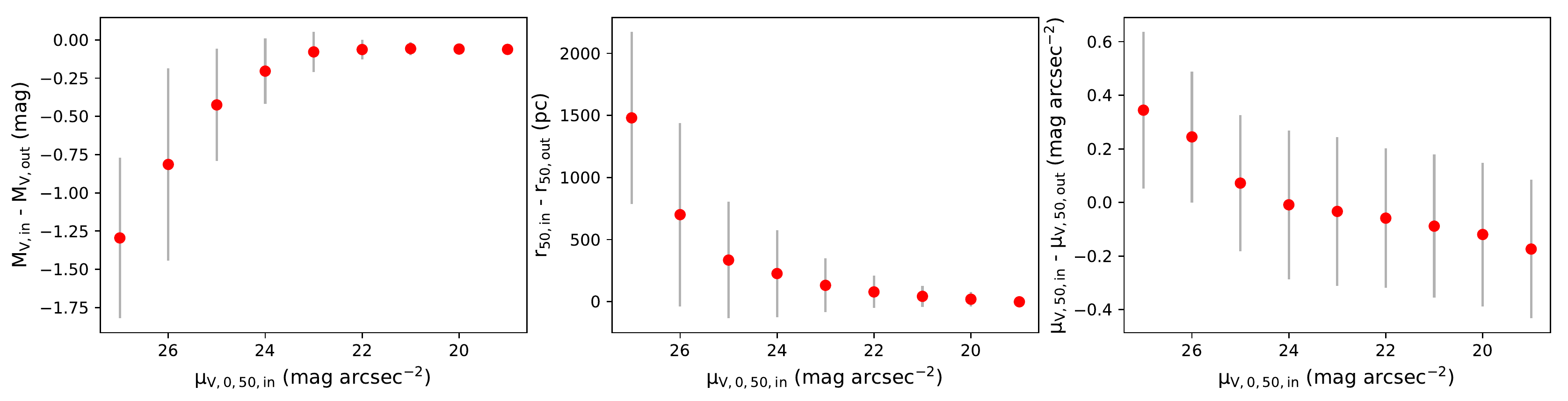}
    \caption{Uncertainty estimate for Petrosian $M_{V,0}$, $r_{50}$, and $\langle\mu_{V,0}\rangle_{50}$ based on a comparison of intrinsic to measured parameters of a set of model galaxies inserted into the {\it WHT} $V-$band data. The red dots correspond to the mean offset between intrinsic (in) and measured (out) parameters at a given surface brightness. The gray error bars indicate the average scatter of the measured parameters in the respective surface brightness bin.}
    \label{fig:figA1}
\end{figure}

Figure~\ref{fig:figA1} shows the mean parameter offsets between intrinsic and measured parameters, as well as the average scatter of the  measured parameters at a given surface brightness. The offsets and scatter in magnitude and half-light radius increase toward fainter surface brightnesses, with the measured magnitudes being, on average, fainter, and the measured half-light radii, on average, smaller than the intrinsic parameters. The measured surface brightnesses are, on average, brighter than the intrinsic values for low surface brightness sources and slightly fainter for high surface brightness sources, while the scatter remains roughly constant. The offsets are likely caused by an underestimate of the intrinsic half-light radius for the low surface brightness sources and a slight overestimate of the intrinsic half-light radius for the high surface brightness sources due to lacking PSF deconvolution.

\section{Color Gradients of the Morphological Subclasses}
\label{sec:secA2}

\begin{figure*}[h!]
	\includegraphics[width=\textwidth]{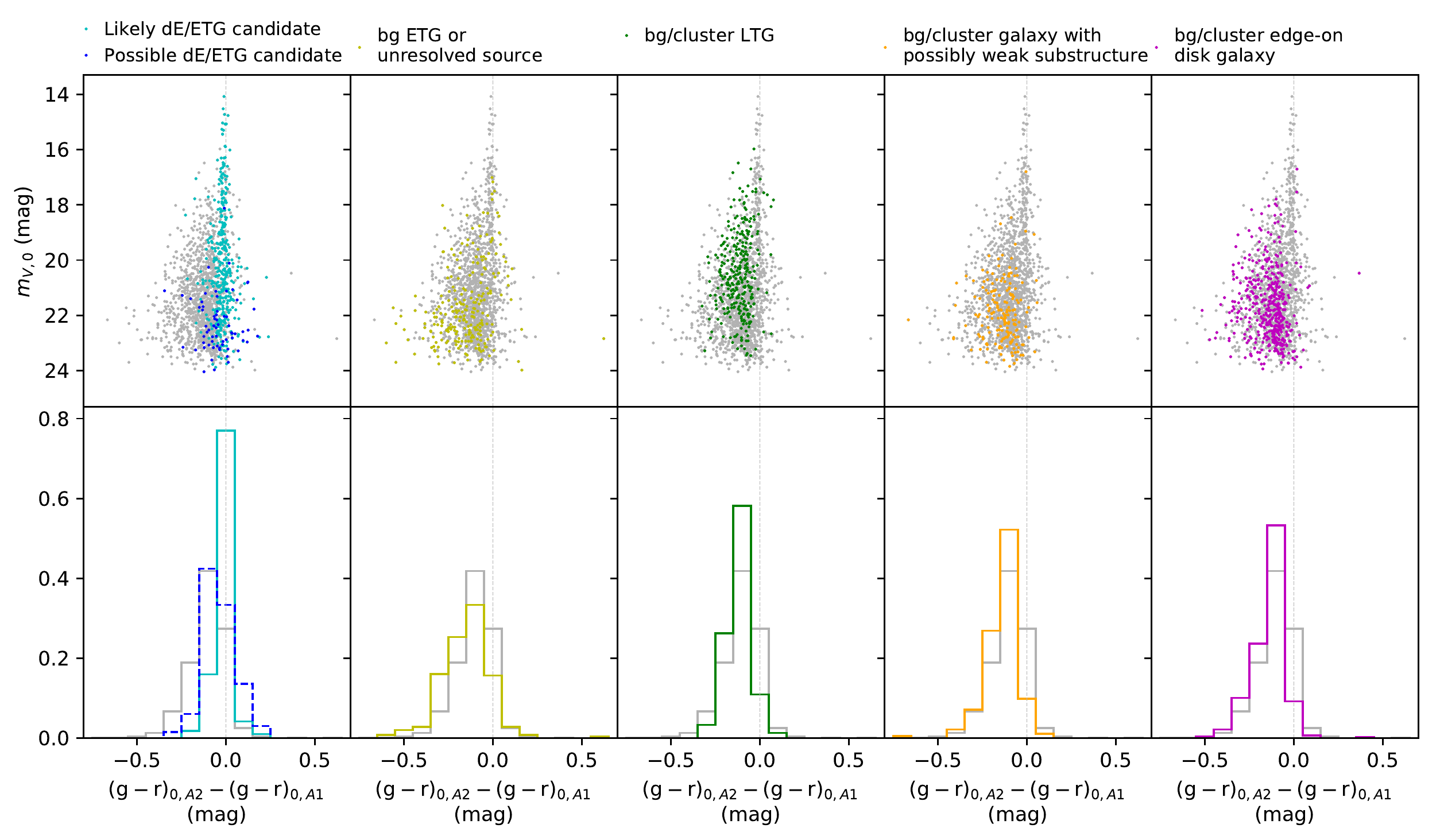}
    \caption{Color gradient versus apparent luminosity (upper panels) and color gradient distribution (lower panels) for the morphological subclasses 1-5, defined in Table~\ref{tab:tab4}. In each panel we highlight one subclass and show the respective other sources as gray symbols. The color gradient is measured as the difference between the aperture 2 (A2) color with $r=2\,r_{50}$ and the aperture 1 (A1) color with $r=1\,r_{50}$. The figure only includes sources with $r_{50} \geq 4$\,pixels.}
    \label{fig:figA2}
\end{figure*}

We measured the color gradient as the difference between the outer aperture color with $r = 2\,r_{50}$ and the inner aperture color with $r = 1\,r_{50}$ for all catalog sources with $r_{50} > 4$\,pixels. We show the color gradient as a function of luminosity and its distribution in Figure~\ref{fig:figA2}.

The vast majority of the dE/ETG candidates show no or only a mild color gradient, with about $80$\,per\,cent having a color gradient less than $\pm0.05$\,mag. This may be expected for faint galaxies with early-type stellar populations \citep{Urich2017}. All other subsamples, in particular those classified as LTGs, galaxies with possible substructure, and edge-on disk galaxies, show a clear offset in their distribution toward a negative gradient, with the majority of the sources being bluer in their outskirts. This may support the point that some of them are star-forming disk galaxies, as discussed in Section~\ref{sec:sec5}. 

\section{Color--Magnitude Diagram: $M_{V,0}$ -- \MakeLowercase{({\it r -- z})$_0$}}
\label{sec:secA3}

\begin{figure*}[h!]
\centering
	\includegraphics[width=0.5\textwidth]{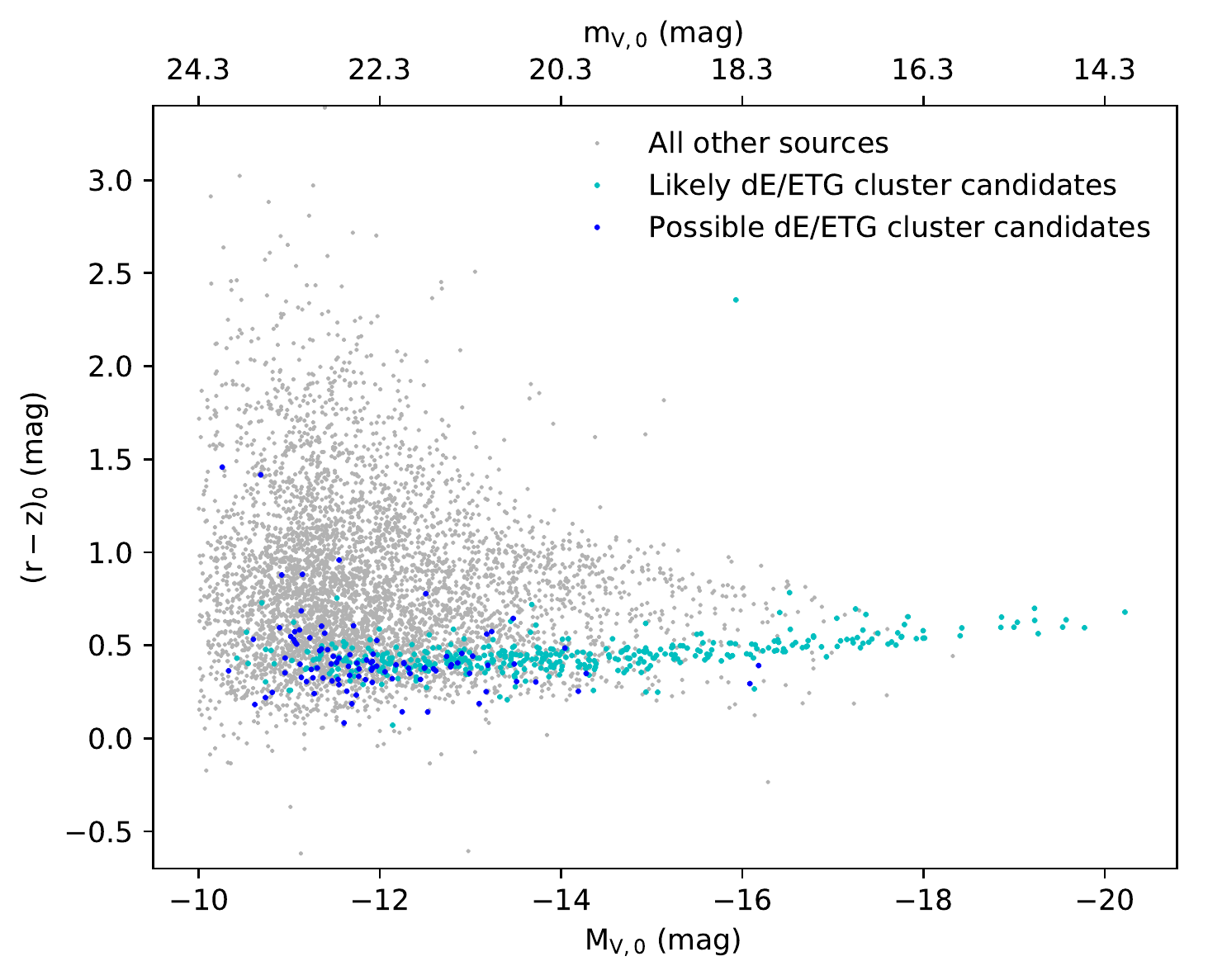}
    \caption{Color--magnitude diagram for the Perseus cluster dE/ETG candidates (highlighted with cyan and blue dots) and other PCC sources (gray dots), including the morphological categories (2) likely background ETGs or sources with unresolved substructure, (3) cluster or background LTGs, (4) cluster or background galaxies with possibly weak substructure, (5) likely cluster or background edge-on disk galaxies. Same as Figure~\ref{fig:fig7} (left panel) but showing $M_{V,0}$ versus ($r-z$)$_0$.}
    \label{fig:figA3}
\end{figure*}

Figure~\ref{fig:figA3} shows the color-magnitude diagram of our morphologically classified sources, highlighting the dE/ETG subsample. Figure~\ref{fig:figA3} is similar to Figure~\ref{fig:fig7}, but shows ($r-z$)$_0$ as a function of magnitude, instead of ($g-r$)$_0$.

\section{{\it HST} Data Comparison}
\label{sec:secA4}

Table~\ref{tab:tabA4} summarizes the {\it HST} data comparison for the Perseus dE/ETG candidates explained in Section~\ref{sec:sec8.1}.

\startlongtable
\begin{deluxetable*}{lll}
\tablecaption{{\it HST} Data Comparison for Perseus dE/ETG Candidates (see Section~\ref{sec:sec8.1}) \label{tab:tabA4}}
\tablecolumns{3}
\tablewidth{0pt}
\tablehead{
\colhead{ID} & \colhead{Comment} & \colhead{{\it HST} Frame}
}
\startdata
PCC 0040 &  dE/ETG-like & 5    \\
PCC 0093 &  dE/ETG-like & 5    \\
PCC 0156 &  dE/ETG-like & 5    \\
PCC 0219 &  dE/ETG-like & 5    \\
PCC 0349 &  dE/ETG-like & 4    \\
PCC 0419 &  dE/ETG-like & 4    \\
PCC 1047 &  dE/ETG-like & 5    \\
PCC 1101 &  dE/ETG-like & 4    \\
PCC 1113 &  dE/ETG-like & 5    \\
PCC 1159 &  dE/ETG-like & 5    \\
PCC 1574 &  dE/ETG-like & 6  \\
PCC 1630 &  dE/ETG-like & 6  \\
PCC 1682 &  dE/ETG-like & 4    \\
PCC 1684 &  dE/ETG-like & 5    \\
PCC 1796 &  dE/ETG-like & 7, 8, 6\\
PCC 1842 &  dE/ETG-like & 7, 8, 6\\
PCC 1870 &  dE/ETG-like & 3    \\
PCC 1884 &  dE/ETG-like & 2    \\
PCC 1918 &  dE/ETG-like & 7    \\
PCC 1925 &  dE/ETG-like & 3    \\
PCC 1944 &  dE/ETG-like & 2    \\
PCC 3414 &  dE/ETG-like & 3    \\
PCC 3421 &  dE/ETG-like & 3    \\
PCC 3444 &  dE/ETG-like & 2    \\
PCC 3529 &  dE/ETG-like & 7    \\
PCC 3538 &  dE/ETG-like & 2    \\
PCC 3558 &  dE/ETG-like & 3    \\
PCC 3613 &  dE/ETG-like & 2    \\
PCC 3623 & Undetected/barely seen  & 7, 8  \\
PCC 3687 &  dE/ETG-like & 7, 8  \\
PCC 3697 &  dE/ETG-like & 7, 8  \\
PCC 3749 &  dE/ETG-like & 3    \\
PCC 3759 &  dE/ETG-like & 1, 8  \\
PCC 3806 &  dE/ETG-like & 1, 7  \\
PCC 3832 &  dE/ETG-like & 7    \\
PCC 3851 &  dE/ETG-like & 3    \\
PCC 3915 &  dE/ETG-like & 1, 8  \\
PCC 3943 &  dE/ETG-like & 1, 8  \\
PCC 3950 &  dE/ETG-like & 9    \\
PCC 3960 &  dE/ETG-like & 9    \\
PCC 4008 & Undetected/barely seen  & 2    \\
PCC 4017 & Undetected/barely seen  & 1    \\
PCC 4060 &  dE/ETG-like & 9    \\
PCC 4375 &  dE/ETG-like & 9    \\
PCC 44 &  dE/ETG-like & 1    \\
PCC 4551 &  dE/ETG-like & 9    \\
PCC 4554 &  dE/ETG-like & 1    \\
PCC 4666 &  dE/ETG-like & 1    \\
PCC 46 & Maybe background & 1    \\
PCC 4719 &  dE/ETG-like & 9    \\
PCC 4750 &  dE/ETG-like & 9    \\
PCC 4752 &  dE/ETG-like & 9    \\
PCC 4780 & Undetected/barely seen  & 11    \\
PCC 4784 &  dE/ETG-like & 11    \\
PCC 4811 &  dE/ETG-like & 11    \\
PCC 4816 &  dE/ETG-like & 10    \\
PCC 4826 &  dE/ETG-like & 10    \\
PCC 4855 &  dE/ETG-like & 10    \\
PCC 4862 &  dE/ETG-like & 10    \\
PCC 4876 &  dE/ETG-like & 10    \\
PCC 4900 &  dE/ETG-like & 10    \\
PCC 4942 & Undetected/barely seen  & 4    \\
PCC 4979 & Undetected/barely seen  & 4    \\
PCC 48 & Undetected/barely seen  & 3    \\
PCC 5047 & Undetected/barely seen  & 3    \\
PCC 5118 &  dE/ETG-like & 7    \\
PCC 5121 &  dE/ETG-like & 9    \\
PCC 5136 & Maybe background & 9    \\
PCC 5163 &  dE/ETG-like & 9    \\
PCC 5305 & Undetected/barely seen  & 11    \\
\enddata
\tablecomments{ID: identifier as given in Table~\ref{tab:tab5}. Comment: visual appearance in the {\it HST} frame specified in the third column. {\it HST} frame: 1 = HST\#J91601010, 2 = HST\#J91603010, 3 = HST\#J91604010, 4 = HST\#J91605010, 5 = HST\#J91606010, 6 = HST\#J9BB01010, 7 = HST\#J9BB02010, 8 = HST\#J9BB04010, 9 = HST\#JDKB13010, 10 = HST\#JDKB12010,  11 = HST\#JDKB13030.}
\end{deluxetable*}





\bibliography{perseus_cat_bibliography,extra_refs} 


\listofchanges

\end{document}